\documentclass[twocolumn,noshowpacs,nofootinbib,10pt]{revtex4-1}

\usepackage{float,xcolor,upgreek,yfonts,tikz}
\usepackage{color}
\usepackage{graphicx}
\usepackage{graphicx,float}\usepackage[all]{xy}
\usepackage{amsmath,upgreek}
 \newcommand{\bltx}{\textcolor{black}}
  \newcommand{\blt}{\textcolor{black}}
\usepackage{amssymb}

\usepackage{alphabeta}
\usepackage{textgreek}

\newcommand{\beq}{\begin{eqnarray}}
\newcommand{\eeq}{\end{eqnarray}}
\newcommand{\be}{\begin{equation}}
\newcommand{\ee}{\end{equation}}

\newcommand{\bea}{\begin{eqnarray}}
\newcommand{\eea}{\end{eqnarray}}

\newcommand{\ba}{\begin{eqnarray}}
\newcommand{\ea}{\end{eqnarray}}
\usepackage[colorlinks,hyperindex,unicode]{hyperref}
\definecolor{green1}{RGB}{0,128,0} 
\hypersetup{hidelinks,backref=true,pagebackref=true,hyperindex=true,colorlinks=true,breaklinks=true,urlcolor= blue}
\hypersetup{%
  colorlinks = true,
  linkcolor  = blue,
  citecolor = cyan,
}
\usepackage{bookmark,textgreek}
\usepackage{hyperref,color,xcolor}
\hypersetup{hidelinks,hyperindex=true,colorlinks=true,breaklinks=true,urlcolor= blue}
\hypersetup{%
  colorlinks = true,
  linkcolor  = blue
}
\newcommand\orcidroldao{{\href{https://orcid.org/0000-0003-3978-532X}{\orcidicon}}}
\newcommand{\orcidicon}{%
	\begin{tikzpicture}
	\draw[lime, fill=lime] (0,0)
		circle [radius=0.16]
		node[white] {{\fontfamily{qag}\selectfont \tiny ID}};
	\draw[white, fill=white] (-0.0625,0.095)
		circle [radius=0.007];
	\end{tikzpicture}	\hspace{-2mm}
}

\def\nn{\nonumber }

\begin{document}
\title{Gravitational decoupling and superfluid stars}

\author{Roldao da Rocha\orcidroldao\!\!}
\affiliation{Center of Mathematics,  Federal University of ABC, 09210-580, Santo Andr\'e, Brazil.}
\email{roldao.rocha@ufabc.edu.br}

\begin{abstract} 
The gravitational decoupling is applied to studying minimal geometric deformed (MGD) compact superfluid stars, in covariant logarithmic scalar gravity on fluid branes. The brane finite tension is shown to provide more realistic values for the asymptotic value of the mass function of MGD superfluid stars, besides constraining the range of the self-interacting scalar field, minimally coupled to gravity. 
Several other physical features of MGD superfluid stars, regulated by the finite brane tension and a decoupling parameter, are derived and discussed, with important corrections to the general-relativistic limit that corroborate to current observational data. 
\end{abstract}

\pacs{04.50.Gh, 04.70.Bw, 11.25.-w}

\keywords{Minimal geometric deformation; gravitational decoupling; fluid branes; brane tension; superfluid stars}

\maketitle
\section{Introduction} 
The investigation of stellar distributions, and black holes of stellar origin, currently comprises one of the most relevant topics in gravitational physics. At extremely strong gravitational fields, gravity can be probed by the framework of Einstein's theory of general relativity (GR) and beyond, also encompassing other models that generalize it. Current observations in the Laser Interferometer Gravitational-Wave Observatory (LIGO) and the Evolved Laser Interferometer Space Antenna However (eLISA) can precisely approach any existing theory or model of gravity, by 
the observation of ring-down phases of binary coalescing black hole mergers \cite{LIGOScientific:2017vwq}. 
The minimal geometric deformation (MGD) procedure and other MGD-decoupling methods have been playing an important role in describing a plethora of compact stellar distributions in a wide range of regimes. The MGD-decoupling also includes models with anisotropy and engender analytical solutions of Einstein's effective field equations \cite{Ovalle:2017fgl,ovalle2007,Casadio:2012rf,darkstars}. In the MGD-decoupling setup, the universe can be modeled by a finite tension fluid brane of codimension-1 embedded in a bulk, with GR occupying the single limit when the fluid brane is rigid, or equivalently, when the brane tension attains an infinite value  \cite{Antoniadis:1998ig,Antoniadis:1990ew}. A running MGD parameter, which is inversely proportional to the fluid brane tension, tunes the modifications of MGD-decoupled models to GR. Refs.  \cite{Casadio:2015jva,Casadio:2016aum,Ovalle:2018ans,Fernandes-Silva:2019fez,daRocha:2017cxu,Fernandes-Silva:2018abr} posed the most precise bounds on parameters of several MGD-decoupled solutions, also using LIGO and eLISA data. 

MGD-decoupling methods represent a realistic algorithm to derive and scrutinize compact stars. The finite brane tension drives the ways to deform the Schwarzschild solution of the Einstein's effective field equations  \cite{Ovalle:2013vna,daRocha:2020jdj,Ovalle:2019qyi,Ovalle:2013xla,Ovalle:2018vmg,Ovalle:2018gic}. 
 The cosmic microwave background anisotropy, observed by the Wilkinson Microwave Anisotropy Probe, rules the fluid brane tension to attain finite values and to vary according to the temperature across a cosmological time scale \cite{Abdalla:2009pg,daRocha:2012pt,Casadio:2013uma}, emulating E\"otv\"os fluid membranes \cite{Gergely:2008jr}. Deriving new solutions of Einstein's effective field equations, that are physically feasible, is not a quite straightforward endeavor, mainly due to the nonlinearity of the coupled system of partial differential field equations. The gravitational MGD-decoupling procedure consists of a technique to decouple gravitational sources in GR \cite{Ovalle:2017fgl}. Applying the MGD-decoupling to the metric tensor of a kernel solution of Einstein's effective field equations, one can realize sources
that are split into a core solution and an ancillary source solution. Applying the gravitational MGD-decoupling algorithm to core solutions engenders new physically realistic solutions, including several sorts of stars and their final collapse into black holes \cite{Ovalle:2007bn,Ovalle:2018umz,Estrada:2019aeh,Gabbanelli:2019txr,Pant:2021eit,daRocha:2020rda,Casadio:2017sze}. MGD-decoupled analog models of gravity were proposed in Refs. 
            \cite{daRocha:2017lqj,Fernandes-Silva:2017nec}. 
           The MGD-decoupling has been also scrutinized and applied in diverse frameworks \cite{Ramos:2021drk,Contreras:2021yxe,Casadio:2019usg,Rincon:2019jal,Ovalle:2019lbs,Tello-Ortiz:2019gcl,Morales:2018urp,Panotopoulos:2018law,Singh:2019ktp,Maurya:2020djz,Arias:2020hwz}, also enclosing the analysis and scrutiny of recently observed anisotropic stars  \cite{Ovalle:2017wqi,Gabbanelli:2018bhs,PerezGraterol:2018eut,Heras:2018cpz,Torres:2019mee,Hensh:2019rtb,Contreras:2019iwm,Sharif:2018toc,Stelea:2018cgm,Maurya:2019hds,Cedeno:2019qkf,Deb:2018ccw,Tello-Ortiz:2021kxg,Tello-Ortiz:2020ydf,Zubair:2020lna,Sharif:2020vvk}. Lower dimensional gravitational decoupling models were reported in Refs. \cite{Contreras:2018vph,Contreras:2019fbk,Contreras:2019mhf,Sharif:2019mzv}. Also, the entropy of entanglement in extended MGD-decoupled anisotropic solutions \cite{Casadio:2015gea} was comprehensively scrutinized in Refs. \cite{daRocha:2019pla,daRocha:2020gee}. Sources of fluid anisotropy in stellar distributions, that include neutron stars, white dwarfs and quark stellar distributions, comprise magnetic
fields and viscosity in extremely dense matter \cite{Contreras:2021xkf}. Extensions of the gravitational action that engenders the geometric MGD-decoupling have been also studied, encompassing Lovelock, Brans--Dicke, $f(R)$, and $f(R,T)$ gravity \cite{Sharif:2021goe,Maurya:2021aio,Sharif:2020lbt,Maurya:2020ebd}, also including hairy black holes \cite{Ovalle:2020kpd}. Using trace and Weyl anomalies, MGD-decoupled solutions were shown to play a prominent role on emulating AdS/CFT on the membrane paradigm of braneworld models \cite{Meert:2020sqv}, whereas the complete geometric deformation was proposed in Ref. 
\cite{Maurya:2020gjw}.

In this work, gravitational decoupled superfluid stars are studied in the covariant logarithmic scalar gravity setup, taking into account strongly-interacting and extremely dense superfluids.  
The Klein--Gordon equation with logarithmic potential was proposed in the 1960s, in the context of QFT \cite{Rosen:1969ay}. Thereafter, several contexts have used it to study baryogenesis, inflationary cosmology, and phase transitions in this context, with density inhomogeneities \cite{Rosen:1969ay,Linde:1992gj}. Since inflationary universe models driven by scalar fields with logarithmic potential shed new light on cosmology \cite{Barrow:1995xb,Rosen:1969ay,Linde:1992gj}, it is now natural to argue how the setup of the Klein--Gordon equation with logarithmic potential, coupled to GR, can be used to study MGD-decoupled superfluid stars.

The recently observed gravitational wave emitted from the binary neutron star merger GW170817 was a landmark in the analysis of the extremely dense matter at strong gravity. As relic stellar distributions, their core has a lower temperature than the transition temperature for the emergence of Bardeen--Cooper--Schrieffer-like pairs. It becomes an appropriate condition to superfluid neutrons states in the elastic inner crust \cite{Datta:2019ueq,Char:2018grw}. Pulsar glitches, the thermal X-ray radiation, and the observation that the Cassiopeia
A neutron star has been fastly cooling down, are unequivocal and prominent observational signatures of superfluid stars in nature \cite{Page:2010aw,Andersson:2018xmu,Yu:2017cxe,Leinson:2014cja}. 
We will analyze the finite brane tension influence on the superfluid compact stars, in the gravitational MGD-decoupling setup. 
 It is indeed natural to ascertain, in a realistic model that complies with recent observational data, how the finite tension of the brane can drive new gravitational MGD-decoupled solutions of the Einstein--Klein--Gordon coupled system with logarithmic scalar potential, that encodes compact strongly-interacting and extremely dense superfluids. 
 
 This work is structured in the following way: Sec. \ref{MGD} is dedicated to appraise the MGD as a particular case of the gravitational decoupling algorithm and to describe stars on fluid branes with finite tension. Sec. \ref{decplled} is devoted to scrutinizing the MGD-decoupling in covariant logarithmic scalar gravity, describing superfluid stars. The resulting Einstein--Klein--Gordon system with self-interacting logarithmic potential is then numerically solved, discussed, and analyzed in detail. The finite brane tension setup, 
complying with current observational data, is shown to improve the Schwarzschild GR analysis, paving new ways to describe realistic models for superfluid stars. Sect. \ref{4} is reserved for drawing the concluding remarks.

\section{The MGD and MGD-decoupling protocols}
\label{MGD}
 The MGD algorithm has been comprehensively used to construct new solutions of Einstein's effective field equations, also encoding improvements to classical general-relativistic solutions due to a fluid brane scenario \cite{ovalle2007,Ovalle:2013xla,darkstars,Abdalla:2009pg}. In the MGD-decoupling setup, the most accurate brane tension bound $\gamma \gtrapprox 2.8131\times10^{-6} \;{\rm GeV}$ \cite{Fernandes-Silva:2019fez} was predicted. Since the 4-dimensional fluid brane that describes our universe is immersed into a bulk, the Gauss--Codazzi--Mainardi equations compels the bulk Riemann tensor to split into two geometrically distinct terms. The first one is the Riemann tensor on the brane and the second term consists of a sum of quadratic expressions involving the extrinsic curvature, that encrypts the part of the curvature due to the embedding of the brane in the bulk. The effective Einstein's equations on the brane read 
\begin{equation}
\label{5d4d}
R_{\mu\nu} - \frac12\mathsf{R}g_{\mu\nu}
=\Uplambda_4 g_{\mu\nu}+\mathsf{T}_{\mu\nu},
\end{equation} where $\mathsf{R}$ is the Ricci scalar, $8\pi G_4=1$ is adopted, being $G_4$ the Newton's constant on the brane, $R_{\mu\nu}$ represents the Ricci curvature tensor, and $\Uplambda_4$ is the Einstein's cosmological constant on the brane. 
One can split the stress-energy-momentum tensor in (\ref{5d4d}) into \cite{GCGR} 
\beq
\mathsf{T}_{\upsilon\tau}
=
T_{\upsilon\tau}+\mathsf{E}_{\upsilon\tau}+\gamma^{-1} \mathsf{S}_{\upsilon\tau}
+L_{\upsilon\tau}+P_{\upsilon\tau}.\label{tmunu}\eeq 
The first term, $T_{\upsilon\tau}$, is the stress-energy-momentum tensor that represents ordinary and dark matter, also encoding dark energy; the purely electric component of the Weyl tensor, $\mathsf{E}_{\upsilon\tau}$, is non-local and depends on the inverse of the brane tension; the $\mathsf{S}_{\upsilon\tau}$ tensor consists of quadratic combinations of the stress-energy-momentum tensor\footnote{Due to the current cosmological data, cubic (and higher) order terms involving $T_{\upsilon\tau}$ can be neglected.},  	
\begin{eqnarray}\label{smunu}
\!\!\!\mathsf{S}_{\upsilon\tau} \!=\! \frac{T}{3}T_{\upsilon\tau}\!-\!T_{\upsilon \mu}T^\mu_{\ \tau} \!+\! \frac{g_{\upsilon\tau}}{6} \Big[3T_{\mu\sigma}T^{\mu\sigma} - T^2\Big]
\end{eqnarray}
\noindent where $T=g_{\upsilon\tau}T^{\upsilon\tau}$; the tensor $L_{\upsilon\tau}$ dictates the way  
that the fluid brane is immersed and bent into the codimension-1 bulk; and  $P_{\upsilon\tau}$ describes eventual Kaluza--Klein bosons, fermions, and other stringy fields living in the bulk \cite{GCGR,maartens,Antoniadis:2011bi}.
The purely electric component of the Weyl tensor,  
\begin{eqnarray}
\!\!\!\!\!\!\!\!\mathsf{E}_{\upsilon\tau} \!&=&\!-\frac{1}{\gamma}\!\left[ \mathsf{U}\!\left(\!\mathsf{u}_\upsilon \mathsf{u}_\tau \!+\! \frac{1}{3}\mathsf{h}_{\upsilon\tau}\!\right)+ \mathsf{Q}_{(\upsilon} \mathsf{u}_{\tau)}\!+\!\mathsf{P}_{\upsilon\tau}\right], \label{A4}
\end{eqnarray}
\noindent describes a Weyl fluid, where $\mathsf{h}_{\upsilon\tau}$ emulates an induced metric which projects quantities in the orthogonal direction to the Weyl fluid flow velocity field, $\mathsf{u}^\upsilon$. In addition, dark radiation, the non-local anisotropic stress-momentum-tensor, and energy flux, have the following respective expressions \cite{Ovalle:2007bn}
\beq\label{drt}
\mathsf{U}&=&-\gamma\mathsf{E}_{\upsilon\tau} \mathsf{u}^\upsilon \mathsf{u}^\tau,\\\label{eed}
\mathsf{P}_{\upsilon\tau}&=&-\gamma\left(\mathsf{h}_{(\upsilon}^{\;\mu}\mathsf{h}_{\tau)}^{\;\sigma}-\frac13 \mathsf{h}^{\mu\sigma}\mathsf{h}_{\upsilon\tau}\right)\mathsf{E}_{\mu\sigma},\\\label{eee}
\mathsf{Q}_\upsilon &=& -\gamma \mathsf{h}^{\mu\sigma}\mathsf{E}_{\sigma \upsilon}\mathsf{u}^\mu.\eeq 

\blt{Denote hereon by $\ell_p=\sqrt{G_4\hbar/c^3}$ the Planck length and by $G_5$ the 5-dimensional Newton's gravitational constant, that  can be related
to $G_4$ by $G_5 = G_4\ell_p$. A fine tuning involves the 4-dimensional gravitational constant, the bulk cosmological constant, $\Uplambda_5$,
and the brane tension, by \cite{Randall:1999ee}
\be
\Uplambda_4=\frac{\kappa_5^{2}}{2}\Big(\Uplambda_{5}+\frac{1}{6}\kappa_5^{2}\gamma^{2}\Big).\label{fine}\ee
Adopting $\Uplambda_4=0$ yields no exchange of energy between
the bulk and the brane, complying to the conservation law $
\nabla^{\nu}\mathsf{T}_{\mu\nu} = 0$ \cite{Ovalle:2013vna}.
Therefore, Eq. (\ref{fine}), together with the fact that the 4-dimensional coupling constant $\kappa_4 = 8\pi G_4/c^2\approxeq  1.866\times 10^{-26}\,{\rm m}\cdot {\rm kg}{}^{-1}$, and $\kappa_5 = 8\pi G_5/c^2$, are related through the brane tension by $\kappa^2_{4}=\frac{1}{6}\gamma\kappa^4_5$ \cite{maartens}, 
 imply that 
\beq\label{uuup}
\Uplambda_{5}=-\frac{\sqrt{6}}{6}\kappa_4\gamma^{3/2}.
\eeq
what complies to an AdS bulk. 
Besides, the brane tension is related to the 5-dimensional Planck mass by 
$
\gamma\approx \pi r_c\sqrt{{-\Uplambda_5}/{24M_{5}^3}}, 
$ where $
r_c\approx 1\,\mu{\rm m}
$ as stated by the original Randall--Sundrum model \cite{Randall:1999ee}. }

Compact stars are solutions of the Einstein's effective field equations (\ref{5d4d}), with static and spherically symmetric metric 
\begin{equation}\label{abr}
ds^{2} = a(r) dt^{2} - \frac{1}{b(r)}dr^{2} - r^{2}d\Omega^2, 
\end{equation} where $d\Omega^2$ stands for the metric solid angle. 
It implies that the tensor fields in Eqs. (\ref{eed}, \ref{eee}) take a simplified expression, respectively given by $\mathsf{P}_{\upsilon\tau} = \mathsf{P}(\mathsf{u}_\upsilon \mathsf{u}_\tau+ \mathsf{h}_{\upsilon\tau}/3)$ and $\mathsf{Q}_\upsilon = 0$, where $\mathsf{P}=g_{\upsilon\tau}\mathsf{P}^{\upsilon\tau}$. Correspondingly, the brane stress-energy-momentum tensor can be emulated by a perfect fluid, \beq
T_{\upsilon\tau} = (\epsilon + p) \mathsf{u}_\upsilon \mathsf{u}_\tau- pg_{\upsilon\tau},\label{pfc}\eeq with $\mathsf{u}^\upsilon=\sqrt{b(r)}\delta^\upsilon_0$.
The rest of this section is dedicated to show that the MGD and the MGD-decoupling method produce analytical solutions of the  Einstein's effective field equations (\ref{5d4d}, \ref{tmunu}), modeling realistic compact stars \cite{Ovalle:2017fgl,ovalle2007,Casadio:2012rf,darkstars}.

\subsection{MGD algorithm}
\label{MGD1}
The Einstein's effective field equations on the brane, \eqref{5d4d}, denoting by $\mathsf{G}_{\mu\nu} = R_{\mu\nu} - \frac12\mathsf{R}g_{\mu\nu}$ the Einstein's tensor, are equally expressed as
\begin{subequations} 
\beq\label{eqw}
\label{usual} 
\!\!\!\!\!\!\!\!\!\!\!\!\!\!\!\!\!\!b(r)\;&=&1-\frac{1}{r}\int_0^r \!\mathsf{r}^2\!\left[\epsilon(\mathsf{r})
\!+\!\frac{1}{\gamma}\!\!\left(\frac{\epsilon^2(\mathsf{r})}{2}\!+\!\frac{3\mathsf{U}(\mathsf{r})}{16}\right)\!\right]d\mathsf{r},\\
\label{pp}
\!\!\!\!\!\!\!\!\!\!\!\!\!\!\!\!\!\!\!{\mathsf{
P}}{}&=&\frac{\gamma}{6}\left(\mathsf{G}_1^{\,1}-\mathsf{G}_2^{\,2}\right),\\
\label{uu}
\!\!\!\!\!\!\!\!\!\!\!\!\!\!\!\!\!\!\!{\mathsf{U}}&\!=\!&-16\left(\frac{\epsilon^2}{2}\!+\!\frac{16}{3}\epsilon
p\right)\!+\!\gamma\left(2\mathsf{G}_2^{\,2}\!+\!\mathsf{G}_1^{\,1}\right)\!-\!16p \gamma,\\
\label{con1}
\!\!\!\!\!\!\!\!\!\!\!\!\!\!\!\!\!\!\!p^\prime&=&-\frac{a^\prime}{2a}(\epsilon+p),
\eeq
\end{subequations}
where
\beq
\label{g11}\mathsf{G}_1^{\,1}(r)&=&-\frac 1{r^2}+\frac{1}{b(r)}\left( \frac
1{r^2}+\frac1{ra(r)}{\frac{da(r)}{dr}}\right),\nonumber\\
\label{g22}\mathsf{G}_2^{\,2}(r)&\!=\!&\frac{1}{4b(r)}\left[ \frac{2}{a}\frac{d^2a(r)}{dr^2}-\frac{1}{a(r)b(r)}\frac{da(r)}{dr}\frac{db(r)}{dr}\right.\nonumber\\&&\left. \!\!\!\!\!+\frac{1}{a^2(r)}\!\left(\!\frac{da(r)}{dr}\!\right)^2\!+\!\frac{1}{r}\left(\frac{1}{b(r)}\frac{db(r)}{dr}\!-\!\frac{1}{a(r)}\frac{da(r)}{dr}\right)\right].\nonumber
\eeq
 GR is recovered when 
$\gamma\to\infty$.

Eqs. (\ref{eqw} -- \ref{uu}) induce anisotropic effects. The effective density ($\check{\epsilon}$), the radial pressure ($\check{p}_{r}$), and also the tangential pressure ($\check{p}_{\intercal}$), are respectively expressed as \cite{darkstars}
\beq
\check{\epsilon}&=&\epsilon +\strut \displaystyle\frac{1}{\gamma }\left( \frac{\epsilon ^{2}}{2}+\frac{3 \mathsf{U}}{2}\right) \,, \label{efecden}\\
\check{p}_{r}&=&p+\frac{1}{\gamma }\left( \frac{\epsilon ^{2}}{2}+\epsilon \,p+\frac{\mathsf{U}}{2}\right) +\frac{\mathsf{P}}{2\gamma }\,, \label{efecprera}\\
\check{p}_{\intercal}&=&p+\frac{1}{\gamma }\left( \frac{\epsilon ^{2}}{2}+\epsilon p+\frac{\mathsf{U}}{2}\right) -\frac{\mathsf{P}}{2\gamma }.\label{efecpretan}
\eeq
Eq. (\ref{efecden}) illustrates how the MGD stellar density can be split into a sum that involves the GR-like central stellar density and another term regulated by the inverse of the brane tension. This additional term, beyond GR, is a sum of a numerical factor of the dark radiation and a term involving the quadratic GR-like star density. It is worth mentioning that such last terms vanish in the GR case, where the brane tension takes the limit $\gamma\to\infty$.
The coefficient of anisotropy on MGD stars is defined by  
\beq\label{aani}
\Updelta = \check{p}_{r}-\check{p}_{\intercal} = \frac{3\,\mathsf{P}}{2\gamma }.
\eeq

Gravity in the bulk produces the minimal geometric deformation, $\upxi(r)$, in the radial component of the metric,  \begin{eqnarray}
\label{edlrwssg}
b(r)
&=&
\mu(r)+\upxi(r)
\ ,
\end{eqnarray}
where  
\begin{eqnarray}\label{muuu}
\mu(r) = \begin{cases} 1-\frac{1}{r}\int_0^r {\rm r}^2\,\epsilon({\rm r}) d{\rm r}
\,,
&
\quad r\,\leq\,R\,,
\\
1-\frac{2\,{M_0}}{r}
\ ,
&
\quad r>R,
\end{cases}
\end{eqnarray} for $R$ denoting the stellar radius
\beq\label{radius}
R=\frac{\int_0^\infty \mathsf{r}^3\check\epsilon(\mathsf{r})\,d\mathsf{r}}{\int_0^\infty \mathsf{r}^2\check\epsilon(\mathsf{r})\,d\mathsf{r}}.
\eeq
The effective mass function can be alternatively written down as \cite{Casadio:2012rf,Ovalle:2007bn},
 \beq\mathcal{M}(\gamma)=M_0+{\cal O}(\gamma^{-1}),\label{oioi}\eeq where $M_0$ is the GR mass function, owing to the brane tension range bound $\gamma \gtrsim 2.8131\times 10^{-6} \;{\rm GeV}$ \cite{Fernandes-Silva:2019fez}. Therefore Eq. (\ref{muuu}) can be condensed into the Misner--Sharp--Hernandez function, 
\begin{eqnarray}\label{muuu1}
\mu(r) =1-\frac{2\mathcal{M}(r)}{r}.
\end{eqnarray}

\bltx{The solutions of the coupled system (\ref{eqw} -- \ref{con1}) can be derived, when one substitutes Eq. (\ref{uu}) into (\ref{usual}), implying that 
\begin{eqnarray}
\label{e1g}
&&\frac{1}{b}\left(\frac{{4r^2\left(\frac{da}{dr}\right)^2+{4a^2}}}
{r^2\frac{da}{dr}+4ar}-\frac{1}{b}\frac{db}{dr}\right)\nonumber\\
&&=\frac{4a}{r\left(\frac{da}{dr}r+4a\right)}\left[1-{4 a r}{\left(\epsilon-3p-\frac{\epsilon}{\gamma}
(\epsilon+3p)\right)}\right].
\end{eqnarray}
The metric component $b(r)$ thus has the expression
\begin{eqnarray}
\label{primsol}
b(r)\!&=&\!e^{-I}\!\left(\int_0^r\!\!\frac{e^I(\frac{da}{d\mathsf{r}}\mathsf{r}+4a)}{4a\mathsf{r}}\!\left[\frac{2}{\mathsf{r}^2}\!-\!\left(\epsilon\!-\!\frac{\epsilon}{\gamma}\left(\epsilon\!+\!3
p\right)\!-\!3p\right)\right]\!d\mathsf{r}\right.\!\nonumber\\
&&\left.		\qquad\qquad\qquad\qquad\qquad\qquad\qquad+\upbeta(\gamma)\right),
\end{eqnarray}
where the function $\upbeta=\upbeta(\gamma)$ is inversely proportional to the fluid brane tension $\gamma$, with GR regime $\lim_{\gamma\to\infty}\upbeta(\gamma)=0$, } and 
 \beq
\label{I}
\!\!\!\!\!\!\!\!\!\!I(r)=
\int_0^r\frac{\frac{d^2a}{d\mathsf{r}^2}+\left(\frac{da}{d\mathsf{r}}\right)^2(\mathsf{r}^2-1)+4\mathsf{r}\frac{da}{d\mathsf{r}}a+4a^2}
{\mathsf{r}^2\frac{da}{d\mathsf{r}}+{4\mathsf{r}}}\,d\mathsf{r}
\,.
\eeq
The MGD kernel \eqref{edlrwssg} must be compatible to Eq. (\ref{primsol}), yielding  
\begin{eqnarray}\label{kapp}
\!\!\!\!\!\!\!\!\!\!\!\upxi(r)\!=\!{e^{-I(r)}\!\left(\upbeta\!+\!\!\int_0^r\frac{2\mathsf{r}ae^I}{r\frac{da}{d\mathsf{r}}+4a}
\left[\mathsf{L}\!+\!\frac{1}{\gamma}\left(\epsilon^2\!+\!3p\epsilon\right)\right]\right)
d\mathsf{r}},
\end{eqnarray}
where  
\beq
\label{L} \!\!\!\!\!\!\mathsf{L}(r)\;&\!=\!&
\left[\mu\left(\frac{1}{a^2}\frac{d^2a}{dr^2}\!-\!\frac{1}{a^2}\left(\frac{da}{dr}\right)^2\!+\!\frac{2}{ar}\frac{da}{dr}\!+\!\frac{1}{r^2}\right)\!\right.\nonumber\\&&\qquad\quad+\left.\frac{d\mu}{d{r}}\!\left(\frac{1}{2a}\frac{da}{dr}\!+\!\frac{1}{r}\right)-\!\frac{1}{r^2}\right]-3p,
\eeq
impels anisotropic effects onto the brane. The geometric deformation $\upxi(r)$ in the vacuum, $\upxi^\star(r)$, is minimal and is calculated by Eq.~(\ref{kapp}) restricted to $\mathsf{L}(r)=0$ \cite{ovalle2007},
\begin{equation}
\label{def}
\upxi^\star(r)
=\upbeta(\gamma)\,e^{-I}
\ .
\end{equation}
The radial metric component in Eq.~\eqref{edlrwssg} then becomes
\begin{eqnarray}
\label{g11vaccum}
\frac{1}{b(r)}=
{1-\frac{2\,{M_0}}{r}}+\upbeta(\gamma)\,e^{-I}\, ,
\end{eqnarray}
Using Israel--Darmois boundary matching conditions and Eqs. (\ref{muuu}, \ref{muuu1}), in the interior star region, $r<R$, the MGD metric solution reads  
\begin{equation}
ds^{2}
\!=\!
a_{\scalebox{1}{\,-}}(r)dt^{2}
-\frac{1}{1-\frac{2{\mathcal{M}}(r)}{r}}dr^2
-r^{2}d\Omega^2,
\label{mgdmetric}
\end{equation}
with mass function  
\begin{equation}
\label{effecmass}
{\mathcal{M}}(r)
=
{M_0}(r)-\frac{r\upxi^{{\scalebox{0.6}{$\star$}}}(r)}{2}. 
\end{equation}
To match the MGD metric solution in the star interior region to the exterior one, the trace of the anisotropic stress-energy-momentum tensor and
the Weyl dark radiation have the following respective expressions \cite{ovalle2007,darkstars}, 
\ba
\label{pp2}
\,{\mathsf{P}_{\scalebox{0.6}{+}}}{(r)}
&=&
\frac{{4{\mathcal{M}(r)}}{}-3r}{27 \gamma r^4\left(1-\frac{3{\mathcal{M}(r)}}{2r}\right)^{\!2}}\,\upbeta(\gamma),\\
{\mathsf{U}_{\scalebox{0.6}{+}}}{(r)}
&=&
\frac{{\mathcal{M}(r)}}{12 \gamma r^4\left(1-\frac{3{\mathcal{M}(r)}}{2r}\right)^{\!2}}\,\upbeta(\gamma).
\ea
Since $p(r)=0=\epsilon(r)$ for $r>R$, the exterior MGD stellar line element is given by \cite{Casadio:2012rf}
\begin{equation}
\label{genericext}ds^2\!=\!a_{\scalebox{0.6}{+}}(r) dt^2\!-\frac{dr^2}{1\!-\!\frac{2\mathcal{M}(r, \gamma)}{r}\!-\!\upxi^\star(r)}+r^{2}d\Omega^2.
\end{equation}
At the star surface, $r=R$, Israel--Darmois matching conditions imply that \cite{Casadio:2012rf}
\begin{eqnarray}
 \label{ffgeneric1}
\log(a_{\scalebox{1}{\,-}}(R))\;&=&1-\frac{2{\mathcal{M}(R)}}{R}
=\log(a_{\scalebox{0.6}{+}}(R)),\\
 \label{ffgeneric2}
{2\mathcal{M}(R)-2M_0} &=&R\left[\upxi^{\star}_{\scalebox{0.6}{+}}(R)-\upxi^{\scalebox{0.6}{$\star$}}_{\scalebox{0.9}{\,-}}(R)\right].
\end{eqnarray}

The Misner--Sharp--Hernandez, Schwarzschild-like
solution, \beq
a_{\scalebox{.6}{Schw}}(r)=\frac{1}{b_{\scalebox{.6}{Schw}}(r)}
=
1-\frac{2{\mathcal{M}(r)}}{r},\eeq is now put back in Eq.~\eqref{def}, implying the MGD 
\be
\label{defS}
\upxi^\star(r)=
-\frac{4(1-\frac{2\mathcal{M}}{r})}{r\left(2r-{3\mathcal{M}}\right)}\,\upbeta(\gamma).
\end{equation}
Also, at the MGD star surface, the function (\ref{defS}) is negative. Therefore the MGD star horizon, $r_{\scalebox{.61}{MGD}}=2{\mathcal{M}}$, is nearer the star core than the GR Schwarzschild horizon $r_{\scalebox{.6}{Schw}}=2M_0$, as stated by Eq. (\ref{oioi}). 
The explicit formula of the function $\upbeta(\gamma)$ was derived in Ref. \cite{ovalle2007}, 
\begin{equation}
\label{betafinal3}
\upbeta(\gamma)
=\frac{1}{\gamma R}\left(\frac{R-\frac{3{M_0}}{2}}{R-{2\,{M_0}}}\right).
\end{equation}
The MGD star exterior metric components can be then written as \cite{ovalle2007,Casadio:2012rf}
\begin{subequations}
\ba
\label{nu}
\!\!\!\!\!\!a(r)
&=&
1-\frac{2\mathcal{M}}{r}
\ ,
\\
\!\!\!\!\!\!b(r)
&=&
\left(1+\frac{2\mathfrak{l}}{2r-{3\mathcal{M}}}\right)\left(1-\frac{2\mathcal{M}}{r}\right),
\label{mu}
\ea
\end{subequations} 
where 
\begin{equation}
\label{Lk}
\mathfrak{l}=\mathfrak{l}(\gamma)=
\frac{1}{\gamma}
\left(\frac{2R-3M_0}{2R-4M_0}\right)\frac{2R-3\mathcal{M}}{2R-4\mathcal{M}}
\end{equation} 
is the brane tension-dependent MGD parameter. 
The GR limit of a rigid brane, $\gamma\to\infty$, thus recovers the Schwarzschild metric. Using Solar system classical tests of GR, the lower bound range $
|\,\mathfrak{l}\,|\lesssim 6.261\times 10^{-4}\,{\rm m}$ 
 was established \cite{Casadio:2015jva}.  
Also, strong gravitational lensing effects applied to Sagittarius $A^{\scalebox{0.6}{$\star$}}$ set up the bound range $|\mathfrak{l}|\lesssim6.370\times 10^{-2}\ {\rm m}$ \cite{Cavalcanti:2016mbe}. In the limit ${{\mathfrak{l}}}\to 0$, corresponding to a rigid fluid brane of infinite tension, the GR Schwarzschild results are then recovered. 

\subsection{MGD-decoupling method}
\label{MGD2}
The gravitational decoupling method states that the Einstein's effective field equations can be written as \cite{Ovalle:2017fgl}
\begin{equation}\label{eins4}
R_{\mu\nu}-\frac{1}{2}\mathsf{R}g_{\mu\nu}=T_{\mu\nu}^{\scalebox{.6}{decoupled}},
\end{equation}
In \eqref{eins4}, the MGD-decoupled stress-energy-momentum tensor reads 
\begin{equation}\label{eins5}
{T}_{\mu\nu}^{\scalebox{.6}{decoupled}}={T}_{\mu\nu}^{\scalebox{.55}{EFE}}+{T}_{\mu\nu}^{\,\scalebox{.7}{s}}, \end{equation}
for \begin{equation}\label{eins6}
{T}_{\mu\nu}^{\scalebox{.6}{EFE}}=\left(\varepsilon + \mathtt{p}\right)u_{\mu}u_{\nu}-\mathtt{p}g_{\mu\nu},
\end{equation}
 representing the stress-energy-momentum tensor of a core solution of Einstein's effective field equations (EFEs), whereas the term ${T}_{\mu\nu}^{\,\scalebox{.7}{s}}$ evokes a source of anisotropy regulated by a coupling parameter, $\alpha$, as 
\begin{equation}\label{eins7}
{T}_{\mu\nu}^{\,\scalebox{.7}{s}}=\alpha \Uptheta_{\mu\nu},
\end{equation} and $\Uptheta_{\mu\nu}$ denotes ancillary sources of gravity, coupled to the perfect fluid described by ${T}_{\mu\nu}^{\scalebox{.6}{EFE}}$. 

The MGD-decoupling procedure can be described by considering the metric 
\begin{equation}\label{abr2}
ds^{2} = e^{\eta(r)} dt^{2} - e^{-\varsigma(r)}dr^{2} - r^2d\Omega^2, 
\end{equation} 
satisfying Einstein's effective field equations \eqref{eins4}, that can be written as the coupled system
\beq\label{eins11}
\!\!\!\!\!\!\!\!\!\!\!\!\varepsilon+ \alpha \Uptheta_{0}^{\,0}\;&=&\frac{1}{r^2}+e^{-\varsigma}\left(\frac{1}{r}\frac{d\varsigma}{dr}-\frac{1}{r^2}\right) \ ,
\\
\label{eins12}
\!\!\!\!\!\!\!\!\!\!\!\! \alpha \Uptheta_{1}^{\,1}-\mathtt{p}\;&=&\frac{1}{r^2}+e^{-\varsigma}\left(\frac{1}{r}\frac{d\eta}{dr}-\frac{1}{r^2}\right) \ ,
\\
\label{eins13}
\!\!\!\!\!\!\!\!\!\!\!\!\!\alpha \Uptheta_{2}^{\,2}-\mathtt{p}\;&\!=\!&\!\frac{e^{-\varsigma}}{4}\! \left[\!\frac{d\varsigma}{dr}\frac{d\eta}{dr}\!-\!2\frac{d^2\eta}{dr^2}\!-\!\left(\frac{d\eta}{dr}\right)^2\!\right.\nonumber\\&&\left.\qquad\qquad\qquad+\frac{2}{r}\left(\frac{d\eta}{dr}\!-\!\frac{d\varsigma}{dr}\right)\right].
\eeq
The conservation law derived out the Bianchi identity, \begin{equation}\label{eins9}
\nabla^{\nu}{T}^{\scalebox{.6}{decoupled}}_{\mu\nu} = 0, 
\end{equation}
can be also expressed as
\beq\label{eins14}
&&\frac{d\mathtt{p}}{dr}+\frac{1}{2}\frac{d\eta}{dr}\left(\varepsilon+\mathtt{p}\right)-\alpha \frac{d\Uptheta_{1}^{\,1}}{dr}+\frac{2\alpha}{r}\left(\Uptheta_{2}^{\,2}-\Uptheta_{1}^{\,1}\right)\nonumber\\&&\qquad\qquad\qquad\qquad+\frac{\alpha}{2}\frac{d\eta}{dr}\left(\Uptheta_{0}^{\,0}+\Uptheta_{1}^{\,1}\right)=0.
\eeq
Moreover, the effective density, the tangential and radial pressures, respectively read
\beq\label{eins15}
\varepsilon_{\scalebox{.6}{eff}}&=&\varepsilon+\alpha \Uptheta_{0}^{\,0},\\
\label{eins17}
\mathtt{p}_{\intercal}&=&\mathtt{p}-\alpha \Uptheta_{2}^{\,2},\\
\label{eins16}
\mathtt{p}_{r}&=&\mathtt{p}-\alpha \Uptheta_{1}^{\,1}.
\eeq
The anisotropic parameter is expressed by  
\begin{equation}\label{eins18}
\Delta=p_{\intercal}-p_{r}=\alpha (\Uptheta_{2}^{\,2}-\Uptheta_{1}^{\,1}).
\end{equation}
From the coupled system (\ref{eins11} -- \ref{eins16}), the MGD setup can be finally recovered when one observes Eqs. (\ref{pfc}, \ref{abr2}) and \eqref{eins17}, using the notation in the metric \eqref{abr}, with the identifications
\begin{subequations}
\beq\label{eins21}
\eta(r) &\mapsto& \log[a(r)],\\
\label{eins22}
\varsigma(r) &\mapsto&  \varsigma(r)-\log[\alpha \upxi^{\scalebox{0.6}{$\star$}}(r)],\\
\label{eins23}
\mathtt{p}(r)&\mapsto& p(r),\\
\label{eins24}
\upepsilon(r)&\mapsto& \epsilon(r).
\eeq\end{subequations}

\section{MGD superfluid stars}
\label{decplled}
Long before the experimental confirmation of the Higgs boson stated that scalar fields do exist in nature, fundamental scalar fields have been employed in several setups, including quantum field theory, gravitation, and in describing dark matter, inflation, and dark energy as well. Scalar fields can model realistic matter, emulating hydrodynamical fluids governed by some equation of state \cite{Herdeiro:2014goa}. 
To describe MGD superfluid stars, one can emulate the 
paradigm of Bose--Einstein condensates, considering a MGD model coupled to a self-interacting scalar field with covariant logarithmic nonlinearity that arises from dense superfluid models. 
In the GR setup, the Einstein--Klein--Gordon coupled system of field equations has no static solutions for real scalar fields. As GR is a limiting case of the MGD solutions, for $\gamma\to\infty$, it is natural to assume a spherically symmetric minimally-coupled, self-interacting, complex scalar field $\upphi$ hereon. 
The Einstein--Klein--Gordon system is defined by the Lagrangian  
\be\label{lagrange1}
{\cal L} = \mathsf{R}{} -
   g_{\mu\nu}\nabla^\mu \upphi^{\scalebox{0.6}{$\star$}} \nabla^\nu
    \upphi - 
V(\upphi, \upphi^{\scalebox{0.6}{$\star$}}). 
\ee
To describe MGD superfluid stars, potential \be\label{potencial}
V (\upphi, \upphi^{\scalebox{0.6}{$\star$}})
= 
- a_1 \left| \upphi \right|^2 \left[  
  \log \left( 
  	\frac{2\left| \upphi \right|^2}{\kappa}
  \right) - 1 
  \right]
,
\ee
is adopted, where $a_1$ and $\kappa$ are coupling parameters \cite{z12eb,Zloshchastiev:2021ncg}. In superfluid Bose--Einstein condensates models, $a_1$ increases linearly as a function of the 
wave-mechanical temperature
taken as  
a thermodynamic conjugate
of 
the Everett-Hirschman's quantum information entropy  
\cite{Zloshchastiev:2021ncg}. 

The Einstein's effective field equations, can be derived from the Lagrangian \eqref{lagrange1}, as
\beq\label{EFE1}
R^\mu_{\nu} - \frac{1}{2} \delta^\mu_{\nu} \mathsf{R} 
&=& 
 T^\mu_{\nu}, \\
\label{EFE2}
	R_{\mu\nu}&=&-\frac{1}{2}\left[\partial_\mu\upphi\partial_\nu\upphi+g_{\mu\nu}V(\upphi)\right]
\eeq
with stress-energy-momentum tensor 
\ba
T^\mu_\nu 
&=& 
g^{\mu\lambda}
\left(
\nabla_{\lambda}\upphi^{\scalebox{0.6}{$\star$}}\nabla_{\nu}\upphi + \nabla_{\lambda}\upphi\nabla_{\nu}\upphi^{\scalebox{0.6}{$\star$}} 
\right)
\nn\\&& \quad
-
\frac{1}{2}
\delta^\mu_\nu
\left(
g^{\lambda\tau}\nabla_{\lambda}\upphi^{\scalebox{0.6}{$\star$}}\,\nabla_{\tau}\upphi
+
2
V (\upphi,\upphi^{\scalebox{0.6}{$\star$}})
\right)
.
\label{EOMs1}
\ea
The Euler--Lagrange equations of motion that govern the complex scalar field,
\begin{equation}
\label{KGeuler}
\frac{\partial{\cal L}}{\partial\upphi^{\scalebox{0.6}{$\star$}}}
-
\nabla_\mu
\left(\frac{\partial{\cal L}}{\partial\,(\partial_\mu\upphi^{\scalebox{0.6}{$\star$}})}\right)
=
0,
\end{equation} 
yield  
\be\label{esfgen}
 g_{\mu\nu}\nabla^\mu\nabla^\nu \upphi + 2 a_1 
\log{\left(\frac{2|\upphi|^2}{\kappa}\right)}
\,
\upphi = 0.
\ee
The Einstein--Klein--Gordon system has the MGD metric (\ref{abr}) as solutions, as long as stationary scalar fields
\beq
	\upphi(r,t) = e^{-i \omega t} \Upphi(r) 
\eeq
are constrained to the field equations (\ref{EFE1}, \ref{EFE2}) and \eqref{esfgen}, that can be rewritten as the coupled system  
\begin{subequations}\beq
\!\!\!\!\!\!\!\!\!\!\!\!\!\!\!\!-\frac{b'}{\mathtt{x}} 
+
\frac{1+b}{\mathtt{x}^2}
= 2
\left[1-\frac{\Upomega^2}{a} - \log \left(\frac{\uplambda ^2}{\kappa}\right)
\right]\uplambda ^2 
- {\uplambda^{\prime2}}{b}
,
\label{eoms11}\\
\!\!\!\!\!\!\!\!\!\!\!\!\!\!\!\!\frac{ba'}{a \mathtt{x}} 
+
\frac{1+b}{\mathtt{x}^2}
= 
-2\left[
1+\frac{\Upomega^2}{a} - \log \left(\frac{\uplambda ^2}{\kappa}\right)
\right]\uplambda ^2 - {\uplambda^{\prime2}}{b}
,
\label{eoms12}\\
\!\!\!\!\!\!\!\!\!\!\!\!\!\!\!\!\uplambda '' \!+\! 
\left(
\frac{2}{\mathtt{x}} \!-\! \frac{b'}{2b} \!+\! \frac{a'}{2a}
\right)\uplambda' \!+\!  
 \frac{\uplambda}{b}
\left[
\frac{\Upomega^2}{a}\!-\!
 \log \left(\frac{\uplambda ^2}{\kappa}\right)
\right]
\!=\! 0,
\label{eoms13}\eeq
\end{subequations}
where one denotes 
\beq 
\!\!\!\!\!\!\!\!\!\!\!\!\!\!\!\mathtt{x} \!=\! \frac{r}{L},\;\;\;\;L\!=\!\frac{1}{\sqrt{a_1}}, \;\;\;\;\;\uplambda=\frac{\Upphi}{\sqrt{2}},\;\;\; \Upomega
=\frac{\omega}{\sqrt{2 a_1}},\eeq and $(\;\;)^\prime \equiv d/d \mathtt{x}$.
Besides, the mass function now reads \be
\mathcal{M}' (\mathtt{x}) = \mathtt{x}^2 
\left\{
\left[1-\frac{\Upomega^2}{a} + \log \left(\frac{\uplambda ^2}{\kappa}\right)
\right]\uplambda^2
-\frac{b\uplambda^{\prime2}}{2} 
\right\}.
\label{eoms14}
\ee
The coupled system of equations (\ref{eoms11} -- \ref{eoms13}) and (\ref{eoms14}) is compatible to the MGD metric (\ref{abr}, \ref{nu}, \ref{mu}) and can be solved to constrain the scalar field $\uplambda$, with Dirichlet and Neumann conditions 
$\uplambda(0)=\uplambda_{0}$, $\uplambda'(0)=0$, and 
$ \mathcal{M}(0)=0$.

\begin{figure}[h!]
\centering
\centering
	\includegraphics[width=7.8cm]{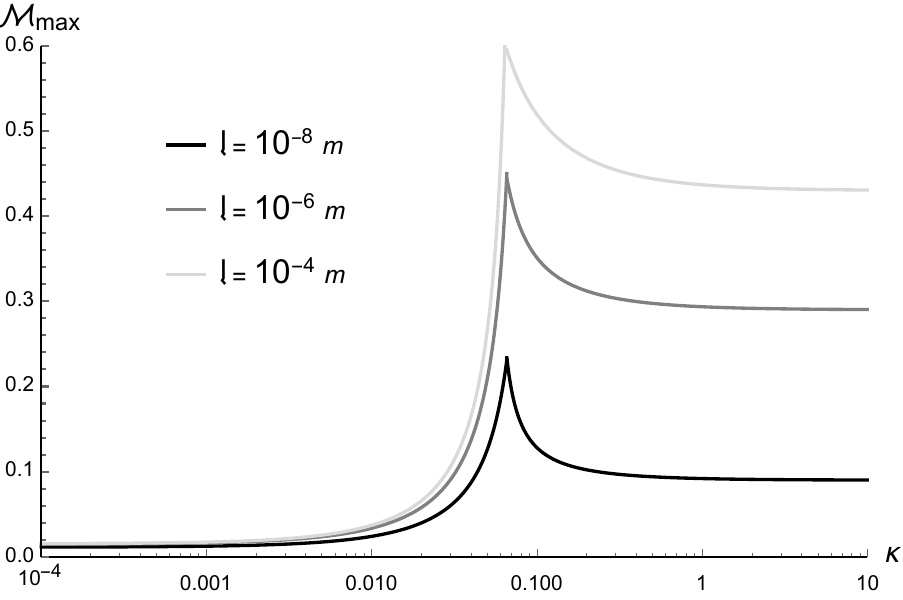}
\caption{ Plot of $\mathcal{M}_\text{max}$ as a function of $\kappa$. The black line shows the case where $\mathfrak{l}=10^{-8}\,{\rm m}$, the grey line corresponds to $\mathfrak{l}=10^{-6}\,{\rm m}$, and the light-grey line regards $\mathfrak{l}=10^{-4}\,{\rm m}$.}
\label{mxx25}
\end{figure}

The value $\kappa_c = 0.06918$ plays a prominent role as a critical value of $\kappa$. We will see in detail that at $\kappa_c$, the behavior of the quantities to be analyzed, in most cases, splits into two distinct cases, namely, for $\kappa \lessgtr \kappa_c$. 
When one analyzes the asymptotic value of the mass function,\beq\label{minfy}
\mathcal{M}_\infty=\lim_{\mathtt{x}\to\infty} \mathcal{M}(\mathtt{x}),\eeq an absolute maximum at
\beq
\mathcal{M}_{\scalebox{.6}{max}}=  
\text{max} \left\{\mathcal{M}_\infty\right\}G_4/L\eeq
is reached at $\kappa = \kappa_c$, where just here we must explicitly adopt the Newton's gravitational constant $G_4$. This profile is presented in Fig. \ref{mxx25}, for different values of the brane tension. Analogously, for all figures that follow, considering the MGD parameter in Eq. (\ref{Lk}) as $\mathfrak{l}=10^{-4}\,{\rm m}$ corresponds to $\gamma \sim 3\times10^{-6} \;{\rm GeV} = 7.3 \times 10^{14}\,
{\rm kg.m^2/s^2}$, whereas $\mathfrak{l}=10^{-6}\,{\rm m}$ and $\mathfrak{l}=10^{-8}\,{\rm m}$ regard, respectively, $\gamma \approx 3\times10^{-4} \;{\rm GeV}$ and $\gamma \approx 3\times10^{-2} \;{\rm GeV}$. Using Eq. (\ref{uuup}), one obtains the value of the 5-dimensional Planck mass corresponding to the values of the brane tension considered here,
\beq
M_5 = \begin{cases}
4.8762\times 10^{-16}\,{\rm kg},&\;\;\; {\rm for}\;\;\;\mathfrak{l}=10^{-4}\,{\rm m},\\
1.0506\times 10^{-15}\,{\rm kg},&\;\;\; {\rm for}\;\;\;\mathfrak{l}=10^{-6}\,{\rm m},\\
2.2637\times 10^{-15}\,{\rm kg},&\;\;\; {\rm for}\;\;\;\mathfrak{l}=10^{-8}\,{\rm m}.
\end{cases}\nonumber
\eeq
Therefore, the analysis hereon takes into account the impact of different values of the brane tension on the physical quantities to be investigated. The higher the value of the brane tension, the more the results approximate the GR limit of an infinitely rigid brane, corresponding to $\gamma\to\infty$.

Fig. \ref{mxx25} shows the maximum asymptotic value of the mass function, $\mathcal{M}_\text{max}$, as a function of the parameter $\kappa$. It illustrates a sharp change of behavior of $\mathcal{M}_{\scalebox{.6}{max}}$ at the critical value $\kappa_c$, when $\kappa$ varies. 
At $\kappa_c$, the quantity $\mathcal{M}_{\scalebox{.6}{max}}$ has a discontinuity of the first derivative $\partial\mathcal{M}_{\scalebox{.6}{max}}/\partial\kappa$ in the graphic, 
irrespectively the value of the brane tension is. For all values of $\kappa$ lower [higher] than $\kappa_c$, the first derivative is non-negative [non-positive]. Besides, Fig. \ref{mxx25} and the numerical analysis involved show that 
\beq
\lim_{\kappa\to\pm\infty}\frac{\partial\mathcal{M}_{\scalebox{.6}{max}}}{\partial\kappa}=0.\eeq  

\begin{figure}[h!]
\centering
	\includegraphics[width=7.8cm]{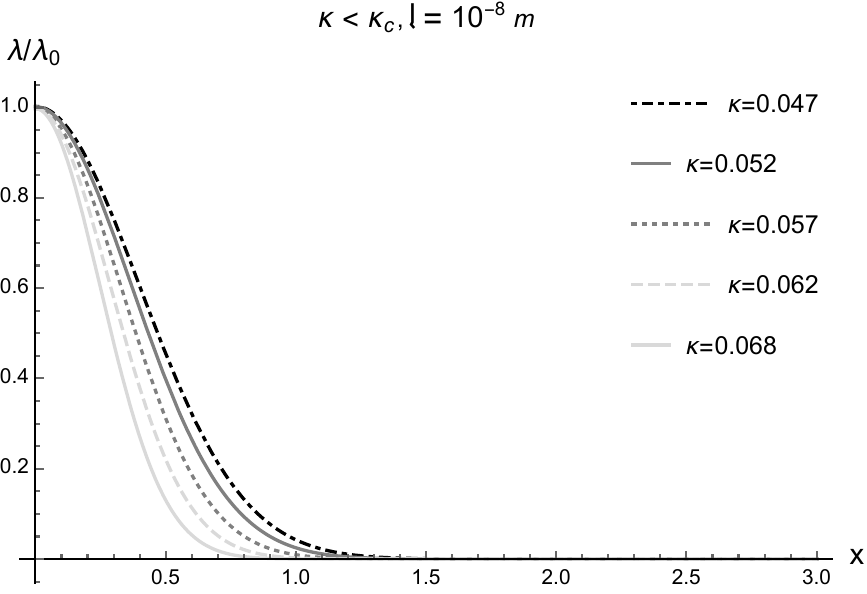}
\caption{Function $\uplambda(\mathtt{x})$ 
of a MGD superfluid star, 
 and $\mathfrak{l}=10^{-8}\,{\rm m}$, for $\kappa<\kappa_c$. The dot-dashed line regards $\kappa=0.047$, the grey line corresponds to  $\kappa=0.052$, the grey dotted line depicts $\kappa=0.057$, the light-grey dashed line illustrates the $\kappa=0.062$ case, whereas $\kappa=0.068$ is shown by the light-grey line.
}
\label{plot1}
\end{figure}

\begin{figure}[h!]
\centering
	\includegraphics[width=7.8cm]{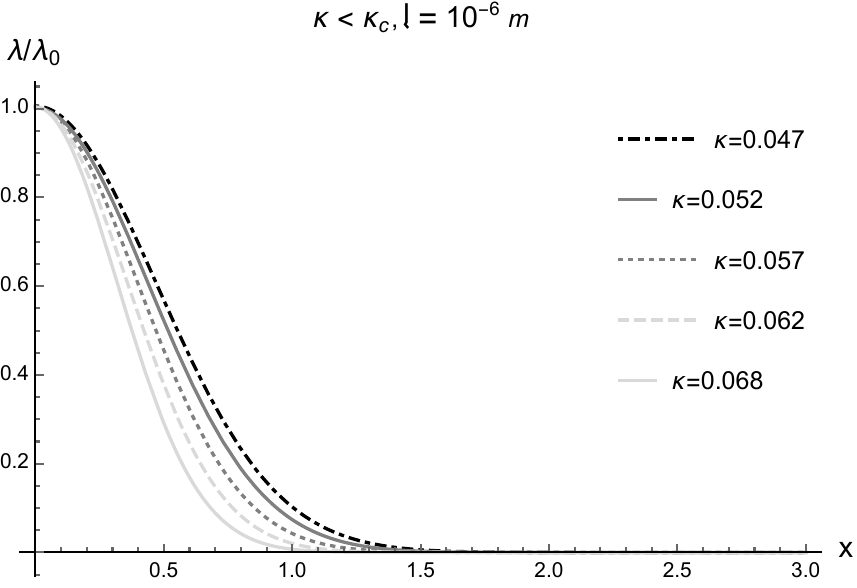}
\caption{Function $\uplambda(\mathtt{x})$ 
of a MGD superfluid star, 
 and $\mathfrak{l}=10^{-6}\,{\rm m}$, for $\kappa<\kappa_c$. The dot-dashed line regards $\kappa=0.047$, the grey line corresponds to  $\kappa=0.052$, the grey dotted line depicts $\kappa=0.057$, the light-grey dashed line illustrates the $\kappa=0.062$ case, whereas $\kappa=0.068$ is shown by the light-grey line.
}
\label{plot2}
\end{figure}

\begin{figure}[h!]
\centering
	\includegraphics[width=7.8cm]{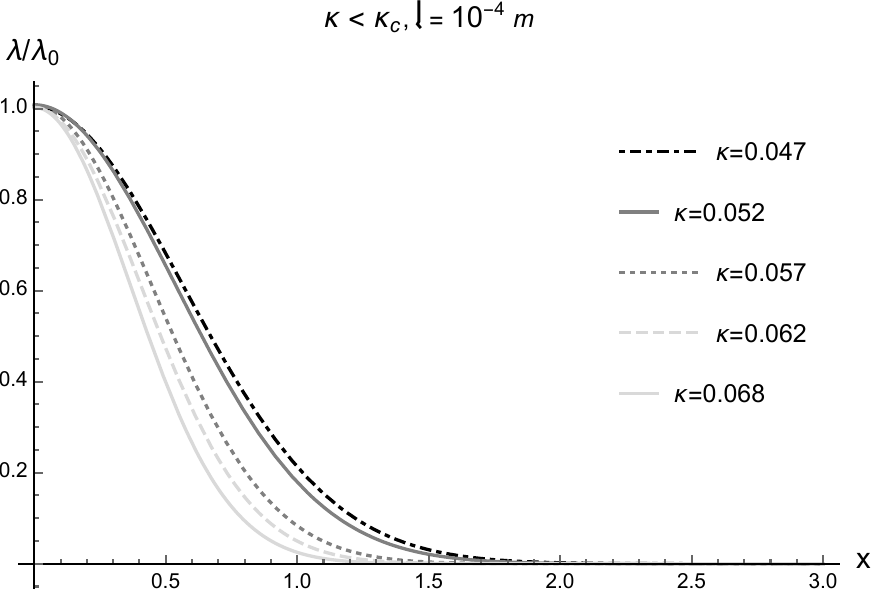}
\caption{Function $\uplambda(\mathtt{x})$ 
of a MGD superfluid star, 
 and $\mathfrak{l}=10^{-4}\,{\rm m}$, for $\kappa<\kappa_c$. The dot-dashed line regards $\kappa=0.047$, the grey line corresponds to  $\kappa=0.052$, the grey dotted line depicts $\kappa=0.057$, the light-grey dashed line illustrates the $\kappa=0.062$ case, whereas $\kappa=0.068$ is shown by the light-grey line.
}
\label{plot3}
\end{figure}

\blt{In what follows the value $\Upomega = 9$ will be employed, following Ref. \cite{Zloshchastiev:2021ncg} that implements the GR case.}
The numerical analysis in Figs. \ref{plot1} -- \ref{plot3} exhibits the profile of the scalar field $\uplambda(\mathtt{x})=\Upphi(\mathtt{x})/\sqrt{2}$ 
of a MGD superfluid star, 
 with respect to the dimensionless radius $\mathtt{x}$, for different values of the brane tension, encoded in the parameter $\mathfrak{l}$, for $\kappa<\kappa_c$. 
Although Figs. \ref{plot1} -- \ref{plot3} illustrate a Gaussian-like profile of the scalar field, the brane tension value clearly alters the rate of variation of $\uplambda(\mathtt{x})$ with respect to $\mathtt{x}$ in each figure. In fact, for each fixed value of $\kappa$, Fig. \ref{plot1} considers the MGD parameter $\mathfrak{l}=10^{-8}\,{\rm m}$, where the scalar field sharply decreases with respect to $\mathtt{x}$. This decrement is attenuated in Fig. \ref{plot2}, where $\mathfrak{l}=10^{-6}\,{\rm m}$ and is blunter in Fig. \ref{plot3}, where $\mathfrak{l}=10^{-4}\,{\rm m}$ is taken into account. It means that lower values of the brane tension can decrease the effective range of the scalar field $\uplambda(\mathtt{x})$. Independently of the brane tension value, the scalar field presents a fast decay, going to zero at the spatial infinity, also exhibiting neither singular points nor nodes.

\begin{figure}[h!]
\centering
	\includegraphics[width=7.8cm]{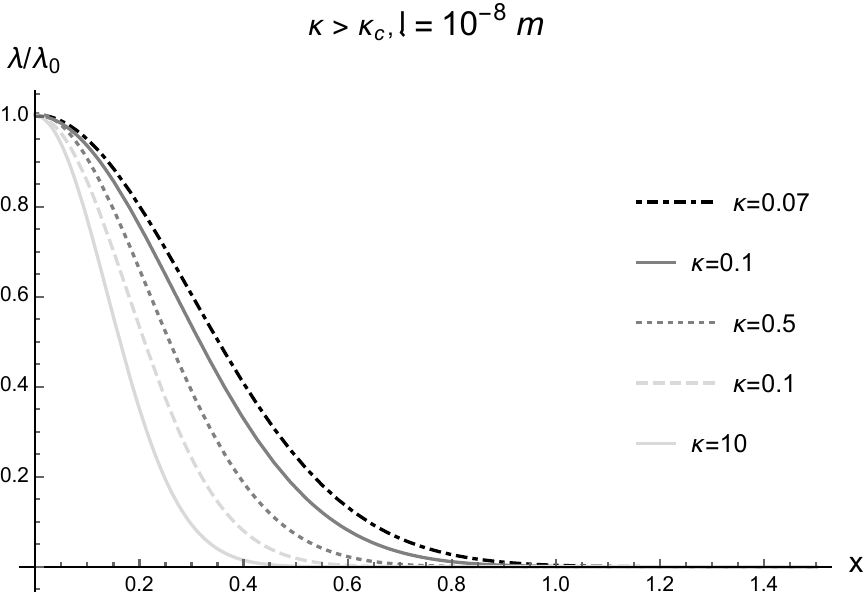}
\caption{Function $\uplambda(\mathtt{x})$ 
of a MGD superfluid star, 
 and $\mathfrak{l}=10^{-8}\,{\rm m}$, for $\kappa>\kappa_c$. The dot-dashed line regards $\kappa=0.07$, the grey line corresponds to  $\kappa=0.1$, the grey dotted line depicts $\kappa=0.5$, the light-grey dashed line illustrates the $\kappa=1$ case, whereas $\kappa=10$ is shown by the light-grey line.
}
\label{plot4}
\end{figure}

\begin{figure}[h!]
\centering
	\includegraphics[width=7.8cm]{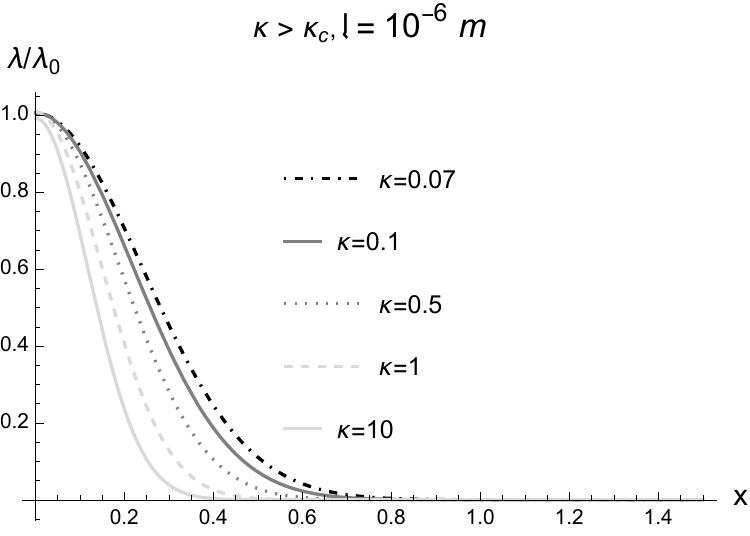}
\caption{Function $\uplambda(\mathtt{x})$ 
of a MGD superfluid star, 
 and $\mathfrak{l}=10^{-6}\,{\rm m}$, for $\kappa>\kappa_c$. The dot-dashed line regards $\kappa=0.07$, the grey line corresponds to  $\kappa=0.1$, the grey dotted line depicts $\kappa=0.5$, the light-grey dashed line illustrates the $\kappa=1$ case, whereas $\kappa=10$ is shown by the light-grey line.
}
\label{plot5}
\end{figure}

\begin{figure}[h!]
\centering
	\includegraphics[width=7.8cm]{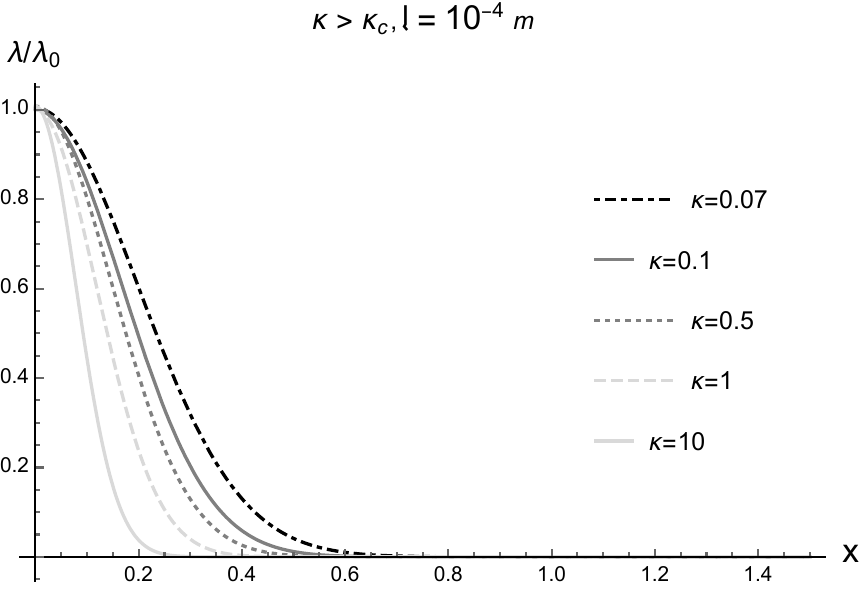}
\caption{Function $\uplambda(\mathtt{x})$ 
of a MGD superfluid star, 
and $\mathfrak{l}=10^{-4}\,{\rm m}$, for $\kappa>\kappa_c$. The dot-dashed line regards $\kappa=0.07$, the grey line corresponds to  $\kappa=0.1$, the grey dotted line depicts $\kappa=0.5$, the light-grey dashed line illustrates the $\kappa=1$ case, whereas $\kappa=10$ is shown by the light-grey line.
}
\label{plot6}
\end{figure}

Figs. \ref{plot4} -- \ref{plot6} also illustrate the scalar field $\uplambda(\mathtt{x})$ 
surrounding a MGD superfluid star, 
 with respect to the dimensionless radius $\mathtt{x}$, for different values of the brane tension, encrypted into the MGD parameter $\mathfrak{l}$. 
Again one can realize a Gaussian-like profile of the scalar field. However, the effect of the brane tension is exactly opposed for values $\kappa>\kappa_c$, when comparing to the respective Figs. \ref{plot1} -- \ref{plot3} that show the range $\kappa<\kappa_c$. In Fig. \ref{plot4}, the MGD parameter  $\mathfrak{l}=10^{-8}\,{\rm m}$ is employed and the scalar field decreases more bluntly with respect to $\mathtt{x}$. This decrement is sharpened in Fig. \ref{plot5}, where $\mathfrak{l}=10^{-6}\,{\rm m}$ and is even sharper in Fig. \ref{plot6}, for $\mathfrak{l}=10^{-4}\,{\rm m}$. It shows that for values of $\kappa$ above the critical value, $\kappa_c$, higher values of the brane tension decrease the reach of the scalar field $\uplambda(\mathtt{x})$ along the radial coordinate, going to zero at the spatial infinity as well. Also, whatever the brane tension value is, the regular scalar field fastly  decays, with no nodes.

\begin{figure}[h!]
\centering
\centering
	\includegraphics[width=8.2cm]{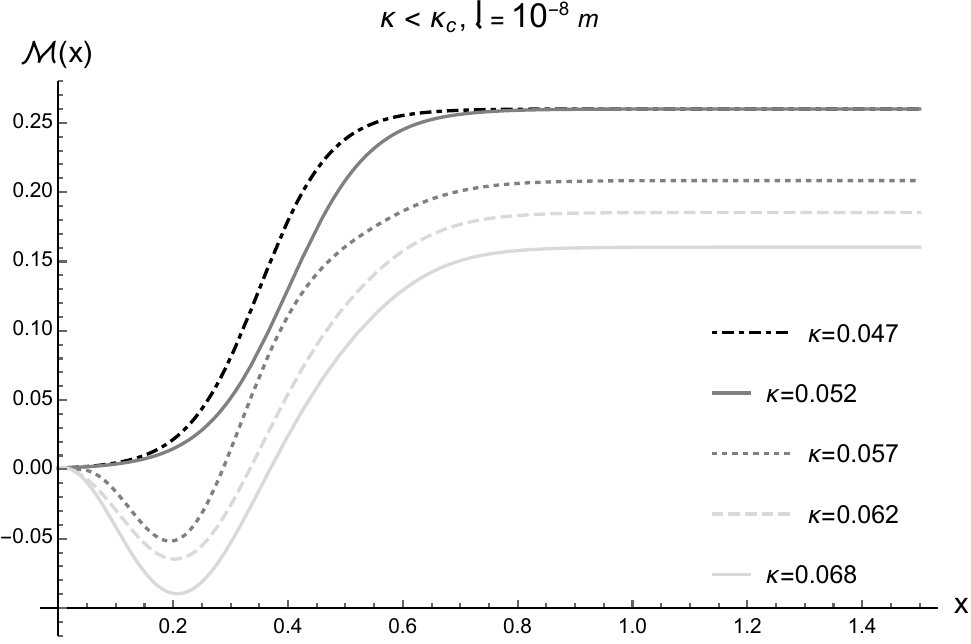}
\caption{
Function $\mathcal{M}(\mathtt{x})$ 
of a MGD superfluid star, for  
 and $\mathfrak{l}=10^{-8}\,{\rm m}$ and $\kappa<\kappa_c$. The dot-dashed line regards $\kappa=0.047$, the grey line corresponds to  $\kappa=0.052$, the grey dotted line depicts $\kappa=0.057$, the light-grey dashed line illustrates the $\kappa=0.062$ case, whereas $\kappa=0.068$ is shown by the light-grey line.}
\label{mxx7}
\end{figure}

\begin{figure}[h!]
\centering
\centering
	\includegraphics[width=8.2cm]{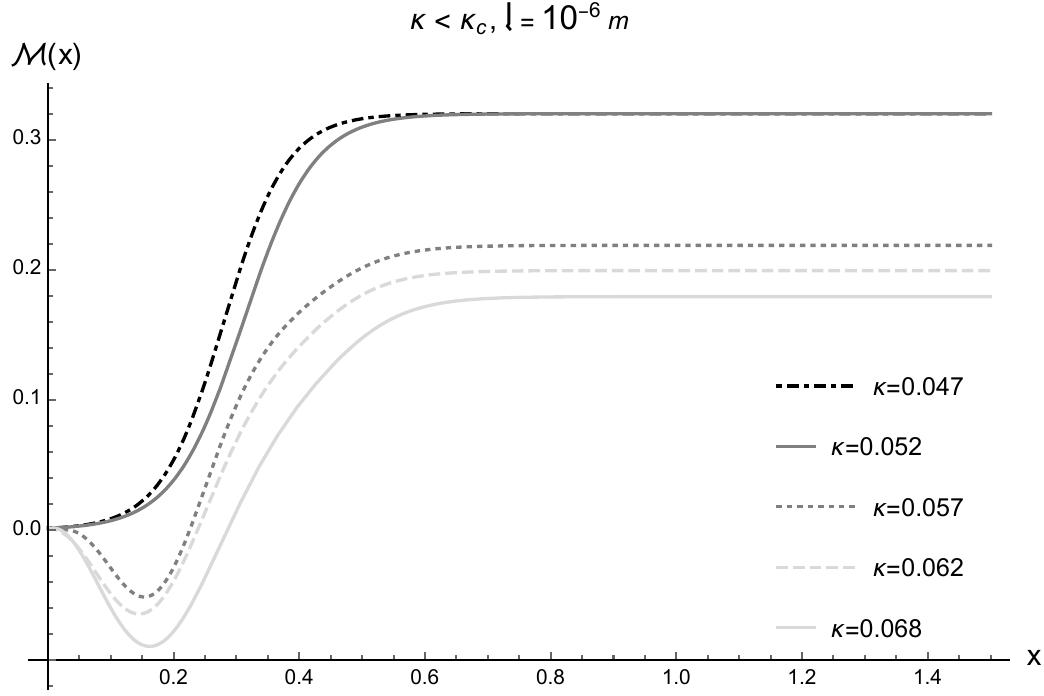}
\caption{
Function $\mathcal{M}(\mathtt{x})$ 
of a MGD superfluid star, for $\mathfrak{l}=10^{-6}\,{\rm m}$ and $\kappa<\kappa_c$. The dot-dashed line regards $\kappa=0.047$, the grey line corresponds to  $\kappa=0.052$, the grey dotted line depicts $\kappa=0.057$, the light-grey dashed line illustrates the $\kappa=0.062$ case, whereas $\kappa=0.068$ is shown by the light-grey line.}
\label{mxx8}
\end{figure}

\begin{figure}[h!]
\centering
\centering
	\includegraphics[width=8.2cm]{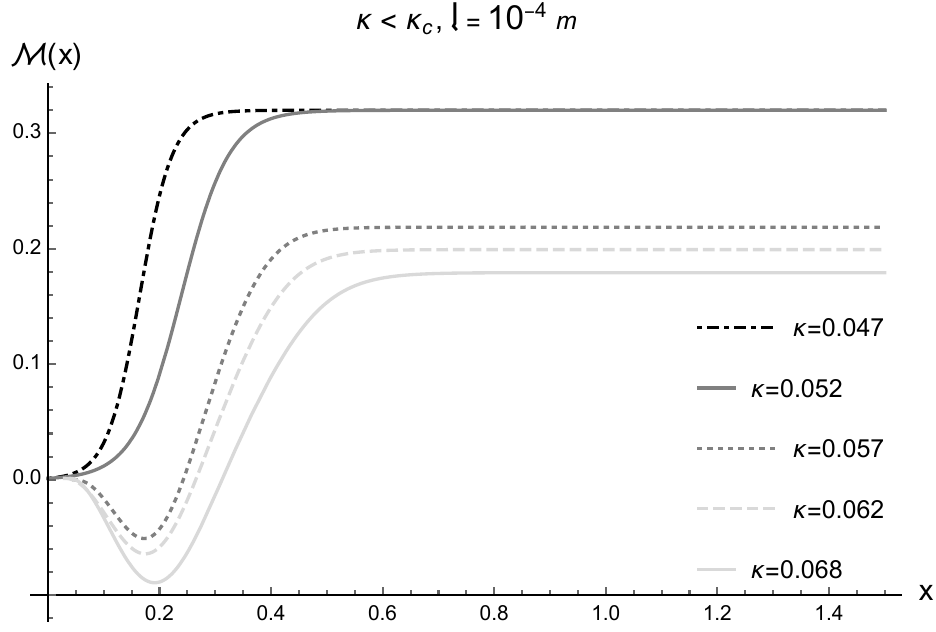}
\caption{
Function $\mathcal{M}(\mathtt{x})$ 
of a MGD superfluid star, 
for $\mathfrak{l}=10^{-6}\,{\rm m}$ and $\kappa<\kappa_c$. The dot-dashed line regards $\kappa=0.047$, the grey line corresponds to  $\kappa=0.052$, the grey dotted line depicts $\kappa=0.057$, the light-grey dashed line illustrates the $\kappa=0.062$ case, whereas $\kappa=0.068$ is shown by the light-grey line.}
\label{mxx9}
\end{figure}

\begin{figure}[h!]
\centering
\centering
	\includegraphics[width=8.2cm]{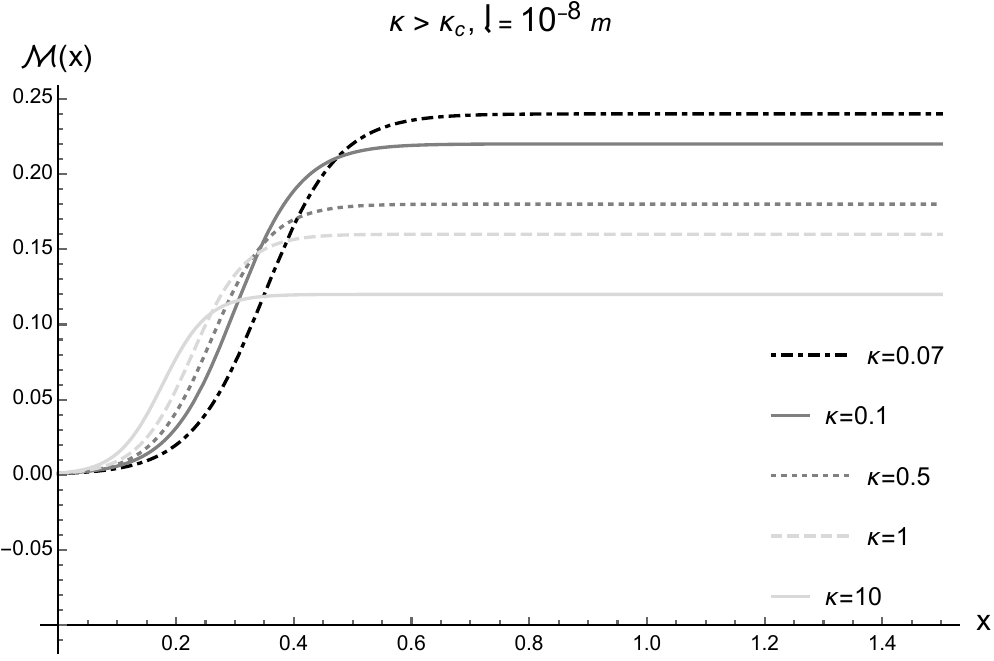}
\caption{
Function $\mathcal{M}(\mathtt{x})$ of a MGD superfluid star, 
 and $\mathfrak{l}=10^{-8}\,{\rm m}$, for $\kappa>\kappa_c$. The dot-dashed line regards $\kappa=0.07$, the grey line corresponds to  $\kappa=0.1$, the grey dotted line depicts $\kappa=0.5$, the light-grey dashed line illustrates the $\kappa=1$ case, whereas $\kappa=10$ is shown by the light-grey line.}
\label{mxx10}
\end{figure}

\begin{figure}[h!]
\centering
\centering
	\includegraphics[width=8.2cm]{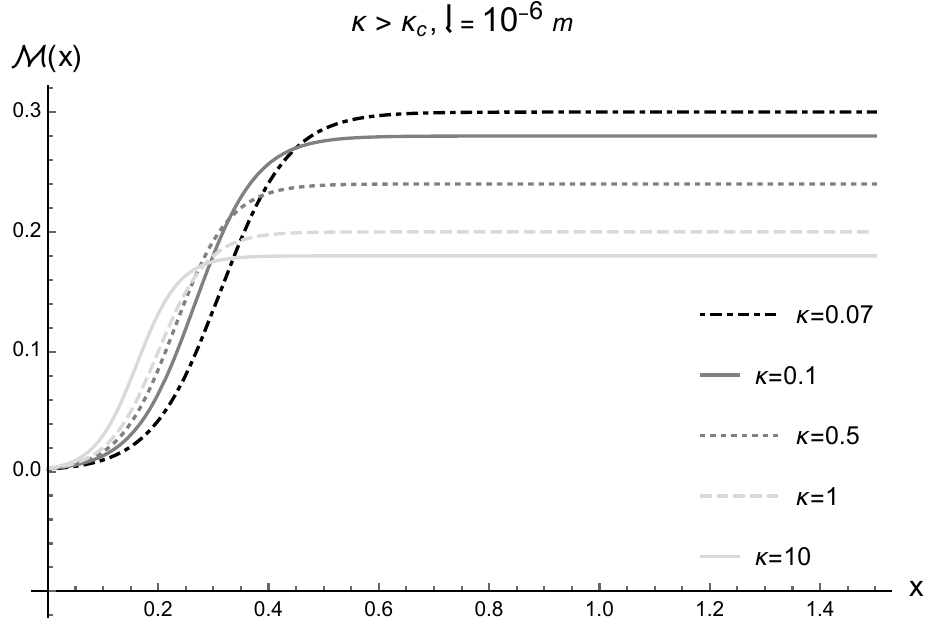}
\caption{
Function $\mathcal{M}(\mathtt{x})$ of a MGD superfluid star, 
 and $\mathfrak{l}=10^{-6}\,{\rm m}$, for $\kappa>\kappa_c$. The dot-dashed line regards $\kappa=0.07$, the grey line corresponds to  $\kappa=0.1$, the grey dotted line depicts $\kappa=0.5$, the light-grey dashed line illustrates the $\kappa=1$ case, whereas $\kappa=10$ is shown by the light-grey line.}
\label{mxx11}
\end{figure}

\begin{figure}[h!]
\centering
\centering
	\includegraphics[width=8.2cm]{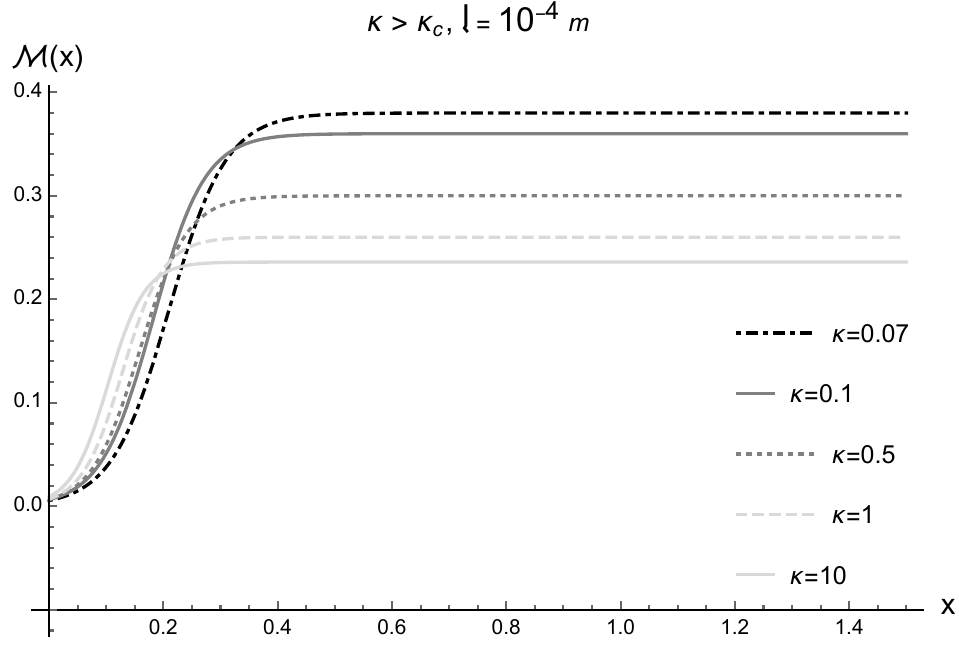}
\caption{
Function $\mathcal{M}(\mathtt{x})$ of a MGD superfluid star, for $\mathfrak{l}=10^{-4}\,{\rm m}$ and $\kappa>\kappa_c$. The dot-dashed line regards $\kappa=0.07$, the grey line corresponds to  $\kappa=0.1$, the grey dotted line depicts $\kappa=0.5$, the light-grey dashed line illustrates the $\kappa=1$ case, whereas $\kappa=10$ is shown by the light-grey line.}
\label{mxx12}
\end{figure}

The profiles of the mass function $\mathcal{M}(\mathtt{x})$ is shown in Figs. \ref{mxx7} -- \ref{mxx12}, respectively for the MGD parameter  $\mathfrak{l}=10^{-8}\,{\rm m}$, $\mathfrak{l}=10^{-6}\,{\rm m}$, and $\mathfrak{l}=10^{-4}\,{\rm m}$, for different values of $\kappa$. 
Fig. \ref{mxx7} plots the mass function $\mathcal{M}(\mathtt{x})$, for $\mathfrak{l}=10^{-8}\,{\rm m}$ and $\kappa<\kappa_c$. Although for all analyzed values of $\kappa$ the mass function $\mathcal{M}(\mathtt{x})$ has an inflection point, for values $\kappa\gtrsim 0.0602$ the mass function presents a negative minimum, which is not physically allowed, as the mass function should be non-negative. Fig. \ref{mxx8} depicts the mass function for $\mathfrak{l}=10^{-6}\,{\rm m}$ and $\kappa<\kappa_c$. Again it also has an inflection point and, for values $\kappa\gtrsim 0.0551$, the mass function presents a negative minimum, 
physically forbidden. The plateau values of the mass function occur for lower values of the dimensionless radius when compared to Fig. \ref{mxx7}. Fig. \ref{mxx9} illustrates the mass function for the value $\mathfrak{l}=10^{-4}\,{\rm m}$ of the MGD parameter and $\kappa<\kappa_c$. For values $\kappa\gtrsim 0.0539$, the mass function does have a negative minimum that is not physically acceptable. The asymptotic values of the mass function happen for even lower values of the dimensionless radius when compared to both Figs. \ref{mxx7} and \ref{mxx8}. Besides, the plateaus of asymptotic values of the mass function increase with increments of the brane tension, for each fixed value of $\kappa$, in Figs. \ref{mxx7} -- \ref{mxx9}.

On the other hand, the issue of a negative-valued mass function does not occur for any value in the range $\kappa> \kappa_c$. Fig. \ref{mxx10} portrays the mass function $\mathcal{M}(\mathtt{x})$, for $\mathfrak{l}=10^{-8}\,{\rm m}$ and $\kappa>\kappa_c$. For all analyzed values of $\kappa$, the higher the value of $\kappa$, the steeper the plot of the mass function is, although the lesser their asymptotic values are. 
Fig. \ref{mxx11} depicts the mass function for $\mathfrak{l}=10^{-6}\,{\rm m}$ and $\kappa>\kappa_c$. Again it changes concavity and the asymptotic values mass function are higher, when compared to the case where $\mathfrak{l}=10^{-8}\,{\rm m}$, in Fig. \ref{mxx10}. Moreover, Fig. \ref{mxx12} shows the mass function when $\mathfrak{l}=10^{-4}\,{\rm m}$. The asymptotic values of the mass function happen for even lower values of the dimensionless radius when compared to both Figs. \ref{mxx10} and \ref{mxx11}, and the higher the value of $\kappa$, the steeper the plot of the mass function is whereas the lesser their asymptotic values are. When comparing Figs. \ref{mxx10} -- \ref{mxx12}, the plateaus of the mass function asymptotic values increase as a function of the brane tension, for each fixed value of $\kappa$.


\begin{figure}[h!]
\centering
\centering
	\includegraphics[width=8.2cm]{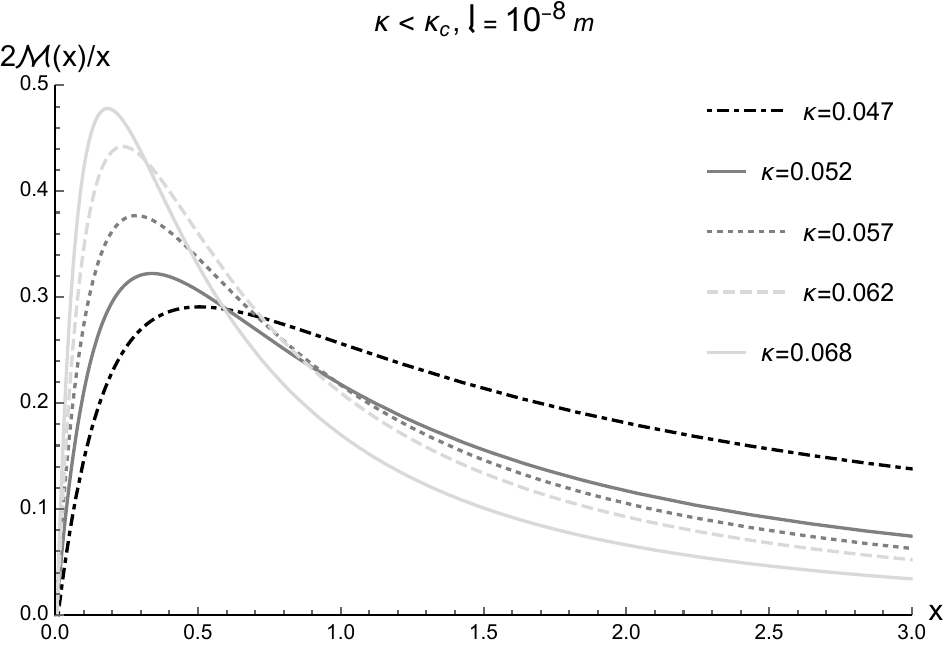}
\caption{
Function $2\mathcal{M}(\mathtt{x})/\mathtt{x}$ of a MGD superfluid star, 
{} and $\mathfrak{l}=10^{-8}\,{\rm m}$, for $\kappa<\kappa_c$. The dot-dashed line regards $\kappa=0.047$, the grey line corresponds to  $\kappa=0.052$, the grey dotted line depicts $\kappa=0.057$, the light-grey dashed line illustrates the $\kappa=0.062$ case, whereas $\kappa=0.068$ is shown by the light-grey line.}
\label{mxx13}
\end{figure}

\begin{figure}[h!]
\centering
\centering
	\includegraphics[width=8.2cm]{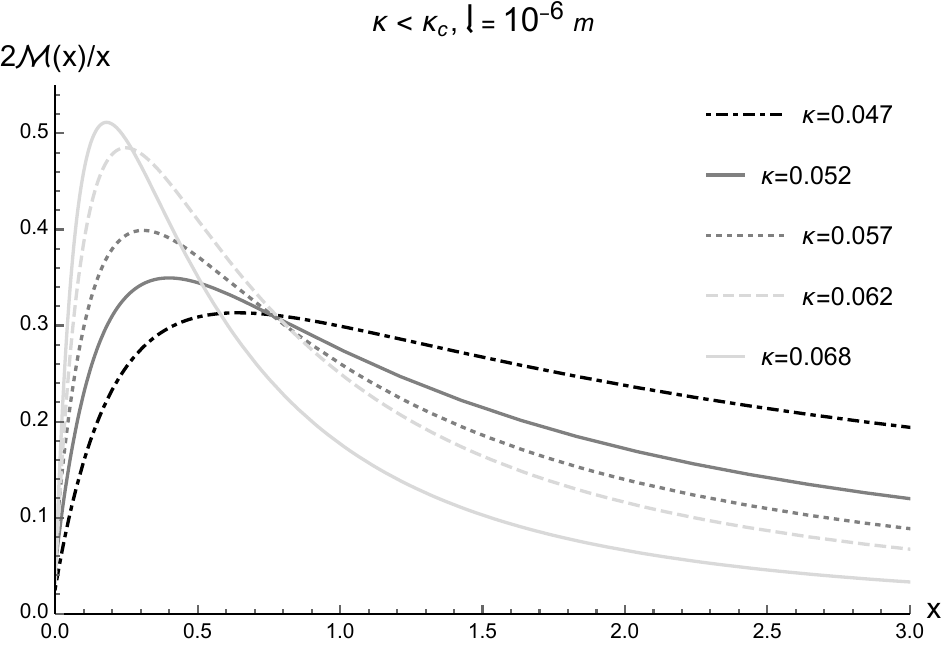}
\caption{
Function $2\mathcal{M}(\mathtt{x})/\mathtt{x}$ of a MGD superfluid star, 
{} and $\mathfrak{l}=10^{-6}\,{\rm m}$, for $\kappa<\kappa_c$. The dot-dashed line regards $\kappa=0.047$, the grey line corresponds to  $\kappa=0.052$, the grey dotted line depicts $\kappa=0.057$, the light-grey dashed line illustrates the $\kappa=0.062$ case, whereas $\kappa=0.068$ is shown by the light-grey line.}
\label{mxx14}
\end{figure}

\begin{figure}[h!]
\centering
\centering
	\includegraphics[width=8.2cm]{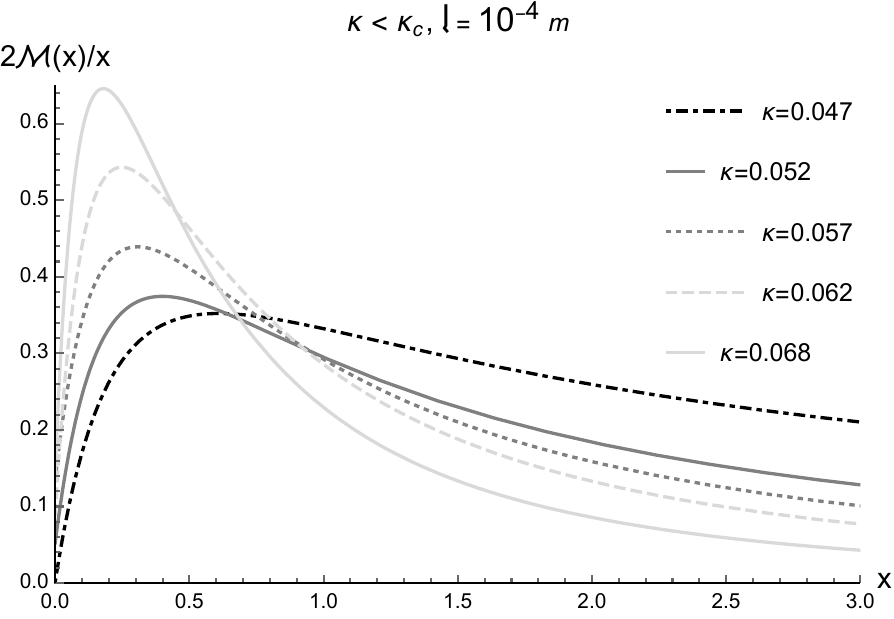}
\caption{
Function $2\mathcal{M}(\mathtt{x})/\mathtt{x}$ of a MGD superfluid star, 
{} and $\mathfrak{l}=10^{-4}\,{\rm m}$, for $\kappa<\kappa_c$. The dot-dashed line regards $\kappa=0.047$, the grey line corresponds to  $\kappa=0.052$, the grey dotted line depicts $\kappa=0.057$, the light-grey dashed line illustrates the $\kappa=0.062$ case, whereas $\kappa=0.068$ is shown by the light-grey line.}
\label{mxx15}
\end{figure}

\begin{figure}[h!]
\centering
\centering
	\includegraphics[width=8.2cm]{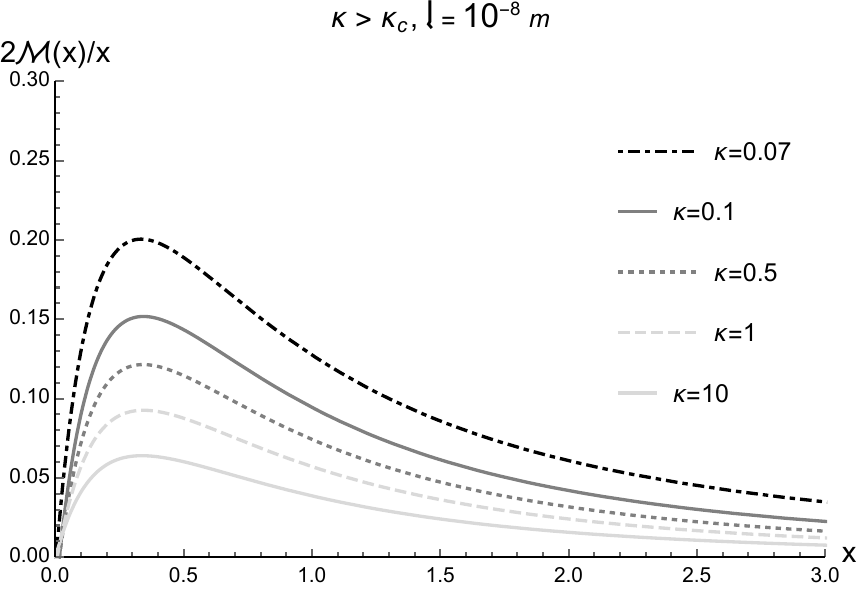}
\caption{
Function $2\mathcal{M}(\mathtt{x})/\mathtt{x}$ of a MGD superfluid star, 
{} and $\mathfrak{l}=10^{-8}\,{\rm m}$, for $\kappa>\kappa_c$. The dot-dashed line regards $\kappa=0.07$, the grey line corresponds to  $\kappa=0.1$, the grey dotted line depicts $\kappa=0.5$, the light-grey dashed line illustrates the $\kappa=1$ case, whereas $\kappa=10$ is shown by the light-grey line.}
\label{mxx16}
\end{figure}

\begin{figure}[H]
\centering
\centering
	\includegraphics[width=8.2cm]{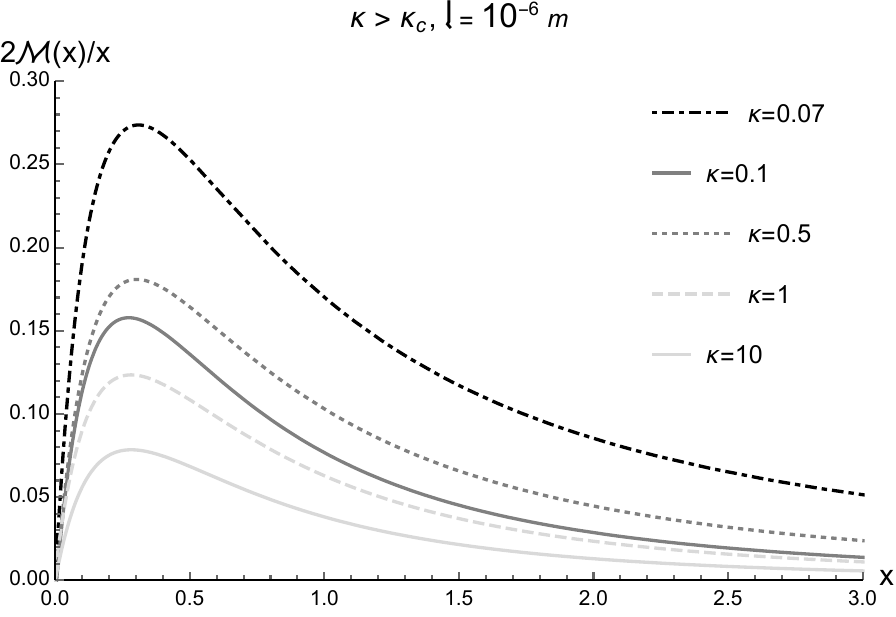}
\caption{
Function $2\mathcal{M}(\mathtt{x})/\mathtt{x}$ of a MGD superfluid star, 
{} and $\mathfrak{l}=10^{-6}\,{\rm m}$, for $\kappa>\kappa_c$. The dot-dashed line regards $\kappa=0.07$, the grey line corresponds to  $\kappa=0.1$, the grey dotted line depicts $\kappa=0.5$, the light-grey dashed line illustrates the $\kappa=1$ case, whereas $\kappa=10$ is shown by the light-grey line.}
\label{mxx17}
\end{figure}

\begin{figure}[H]
\centering
\centering
	\includegraphics[width=8.2cm]{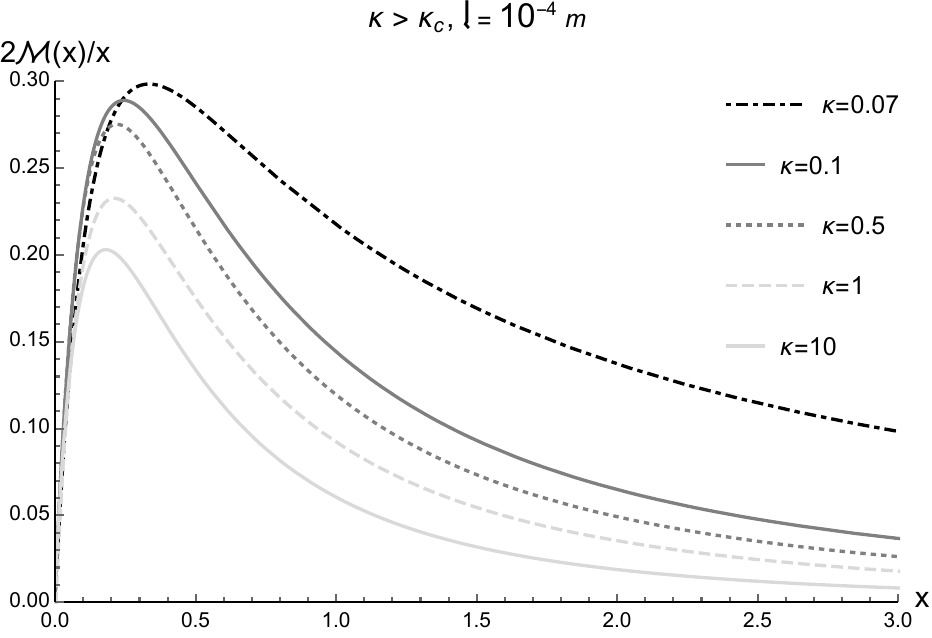}
\caption{
Function $2\mathcal{M}(\mathtt{x})/\mathtt{x}$ of a MGD superfluid star, 
{} and $\mathfrak{l}=10^{-4}\,{\rm m}$, for $\kappa>\kappa_c$. The dot-dashed line regards $\kappa=0.07$, the grey line corresponds to  $\kappa=0.1$, the grey dotted line depicts $\kappa=0.5$, the light-grey dashed line illustrates the $\kappa=1$ case, whereas $\kappa=10$ is shown by the light-grey line.}
\label{mxx18}
\end{figure}

Figs. \ref{mxx13} -- \ref{mxx18} show the relativisticity function ($2\mathcal{M}(\mathtt{x})/\mathtt{x}$) of a MGD superfluid star, for distinct values of $\kappa$. Figs. \ref{mxx13} -- \ref{mxx15} regard the range $\kappa>\kappa_c$, wherein the 
function $2\mathcal{M}(\mathtt{x})/\mathtt{x}$ has a negative second derivative and a steep increment from zero, attains an absolute maximum and smoothly decays, after an inflection point. 
For each fixed value of $\kappa$, the higher the brane tension, the higher the peak of $2\mathcal{M}(\mathtt{x})/\mathtt{x}$ is. This can be realized by the data in Table \ref{scalarmasses1}. Now, for fixed values of the fluid brane tension, the lower the value of $\kappa$, the lower the peak of $2\mathcal{M}(\mathtt{x})/\mathtt{x}$ is and, in general, the higher the value of the radial coordinate where it occurs. Table \ref{scalarmasses1} summarizes the 2-tuples corresponding to the (dimensionless) radial coordinate and the respective peaks of the function $2\mathcal{M}(\mathtt{x})/\mathtt{x}$, in Figs. \ref{mxx13} -- \ref{mxx15}. 
\begin{table}[H]
\begin{center}
\medbreak
-------------- $\left(\left(2\mathcal{M}(\mathtt{x})/\mathtt{x}\right)_{\scalebox{.7}{max}}, \mathtt{x}\right)$, $\;\;\;\kappa<\kappa_c$ --------------\medbreak
\begin{tabular}{||c|c|c|c||}
\hline\hline& \;$\mathfrak{l}=10^{-8}\,{\rm m}$\; & \;$\mathfrak{l}=10^{-6}\,{\rm m}$ \;&\; $\mathfrak{l}=10^{-4}\,{\rm m}$\; \\
    \hline\hline
 $\kappa=0.047$&\; (0.191,\,0.479)\;&\; (0.210,\,0.513)\;&\; (0.211,\,0.648) \;\\ \hline
 $\kappa=0.052$&\; (0.231,\,0.442)\;&\; (0.288,\,0.486)\;&\; (0.270,\,0.544) \;\\ \hline
$\kappa=0.057$&\; (0.318,\,0.377)\;&\; (0.325,\,0.391)\;&\; (0.319,\,0.439) \;\\\hline
$\kappa=0.062$&\; (0.\,340,\,0.320)\;&\; (0.420,\,0.344)\;&\; (0.418,\,0.362) \;\\\hline
$\kappa=0.068$&\; (0.523,\,0.281)\;&\; (0.624,\,0.313)\;&\; (0.641,\,0.346) \;\\\hline
\hline\hline
\end{tabular}
\caption{ Maxima of the function $2\mathcal{M}(\mathtt{x})/\mathtt{x}$, for Figs. \ref{mxx13} -- \ref{mxx15}.} \label{scalarmasses1}
\end{center}
\end{table}
\noindent

Figs. \ref{mxx16} -- \ref{mxx18} depict the cases where $\kappa<\kappa_c$. Analogously, the 
function $2\mathcal{M}(\mathtt{x})/\mathtt{x}$ has a negative second derivative and an abrupt increment from zero, attaining an absolute maximum, and smoothly decays, after an inflection point. Also, for each fixed value of $\kappa$, the higher the brane tension, the higher the peak of $2\mathcal{M}(\mathtt{x})/\mathtt{x}$ is. This can be also noted in the data in Table \ref{scalarmasses2}. But the analogies to the analysis of Figs. \ref{mxx13} -- \ref{mxx15} end here. Contrary to that behavior, for fixed values of the fluid brane tension, the lower the value of $\kappa$, the higher the peak of $2\mathcal{M}(\mathtt{x})/\mathtt{x}$ is, and the higher the value of the radial coordinate where it occurs. Besides, after the peak, the quantity $2\mathcal{M}(\mathtt{x})/\mathtt{x}$ fades away, and the limit
\beq
\lim_{\mathtt{x}\to\infty}\frac{2\mathcal{M}(\mathtt{x})}{\mathtt{x}} = 0
\eeq
holds, irrespectively of the value of the brane tension and $\kappa$, in the numerical analysis underlying Figs. \ref{mxx16} -- \ref{mxx18}. It corroborates to an asymptotically flat brane, consistent with the GR limit of MGD-decoupling.  Despite this general behavior, the rate that that $2\mathcal{M}(\mathtt{x})/\mathtt{x}$ decays varies according to the value of the brane tension.

Table \ref{scalarmasses2} summarizes the 2-tuples corresponding to the (dimensionless) radial coordinate and their respective peaks of the function $2\mathcal{M}(\mathtt{x})/\mathtt{x}$ of a MGD superfluid star, in Figs. \ref{mxx16} -- \ref{mxx18}. 
\begin{table}[h!]
\begin{center}
\medbreak
-------------- $\left(\left(2\mathcal{M}(\mathtt{x})/\mathtt{x}\right)_{\scalebox{.7}{max}}, \mathtt{x}\right)$, $\;\;\;\kappa<\kappa_c$ --------------\medbreak
\begin{tabular}{||c|c|c|c||}
\hline\hline& \;$\mathfrak{l}=10^{-8}\,{\rm m}$\; & \;$\mathfrak{l}=10^{-6}\,{\rm m}$ \;&\; $\mathfrak{l}=10^{-4}\,{\rm m}$\; \\
    \hline\hline
 $\kappa=0.07$&\; (0.340,\,0.202)\;&\; (0.303,\,0.276)\;&\; (0.320,\,0.298) \;\\ \hline
 $\kappa=0.01$&\; (0.334,\,0.153)\;&\; (0.301,\,0.181)\;&\; (0.238,\,0.289) \;\\ \hline
$\kappa=0.5$&\; (0.320,\,0.120)\;&\; (0.266,\,0.157)\;&\; (0.217,\,0.276) \;\\\hline
$\kappa=1$&\; (0.317,\,0.091)\;&\; (0.265,\,0.124)\;&\; (0.201,\,0.234) \;\\\hline
$\kappa=10$&\; (0.305,\,0.062)\;&\; (0.258,\,0.078)\;&\; (0.184,\,0.209) \;\\\hline
\hline\hline
\end{tabular}
\caption{ Maxima of the function $2\mathcal{M}(\mathtt{x})/\mathtt{x}$, for Figs. \ref{mxx16} -- \ref{mxx18}.} \label{scalarmasses2}
\end{center}
\end{table}
\noindent

\begin{figure}[h!]
\centering
\centering
	\includegraphics[width=8.2cm]{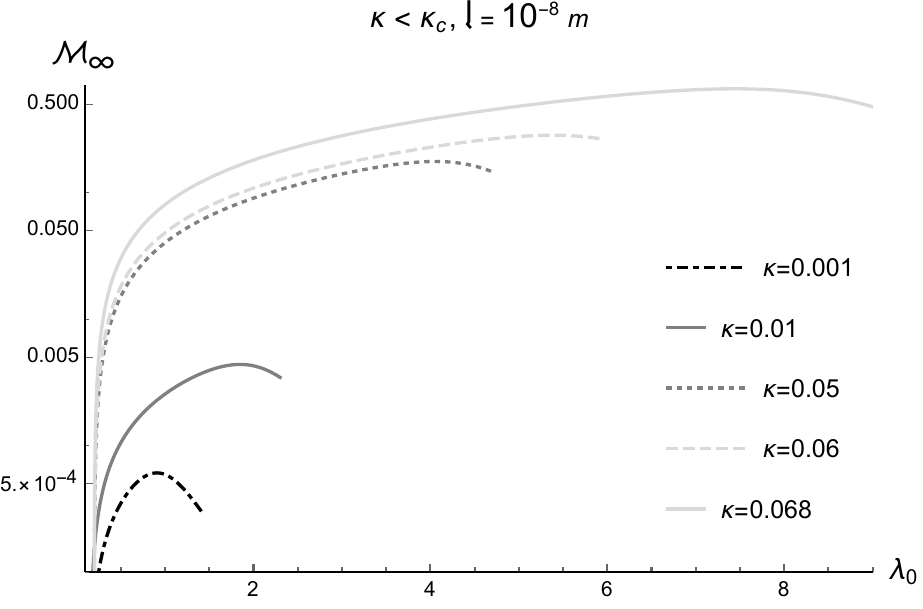}
\caption{
$\mathcal{M}_\infty$ as a function of the central field $\uplambda_{0}$, for $\mathfrak{l}=10^{-8}\,{\rm m}$, for $\kappa<\kappa_c$. The dot-dashed line regards $\kappa=0.001$, the grey line corresponds to  $\kappa=0.01$, the grey dotted line depicts $\kappa=0.05$, the light-grey dashed line illustrates the $\kappa=0.06$ case, whereas $\kappa=0.068$ is shown by the light-grey line.}
\label{mxx19}
\end{figure}

\begin{figure}[h!]
\centering
\centering
	\includegraphics[width=8.2cm]{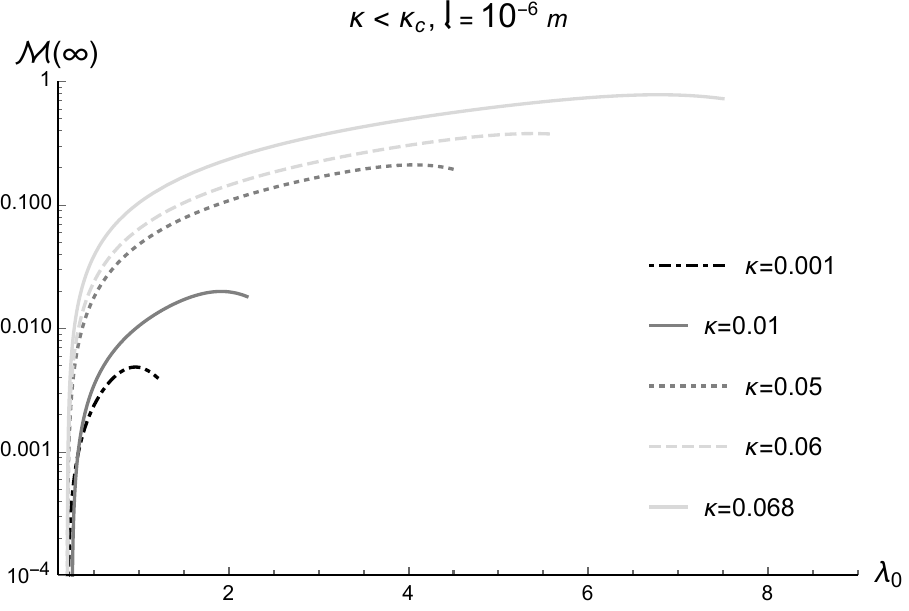}
\caption{
$\mathcal{M}_\infty$ as a function of the central field $\uplambda_{0}$, for $\mathfrak{l}=10^{-6}\,{\rm m}$, for $\kappa<\kappa_c$. The dot-dashed line regards $\kappa=0.001$, the grey line corresponds to  $\kappa=0.01$, the grey dotted line depicts $\kappa=0.05$, the light-grey dashed line illustrates the $\kappa=0.06$ case, whereas $\kappa=0.068$ is shown by the light-grey line.}
\label{mxx20}
\end{figure}

\begin{figure}[h!]
\centering
\centering
	\includegraphics[width=8.2cm]{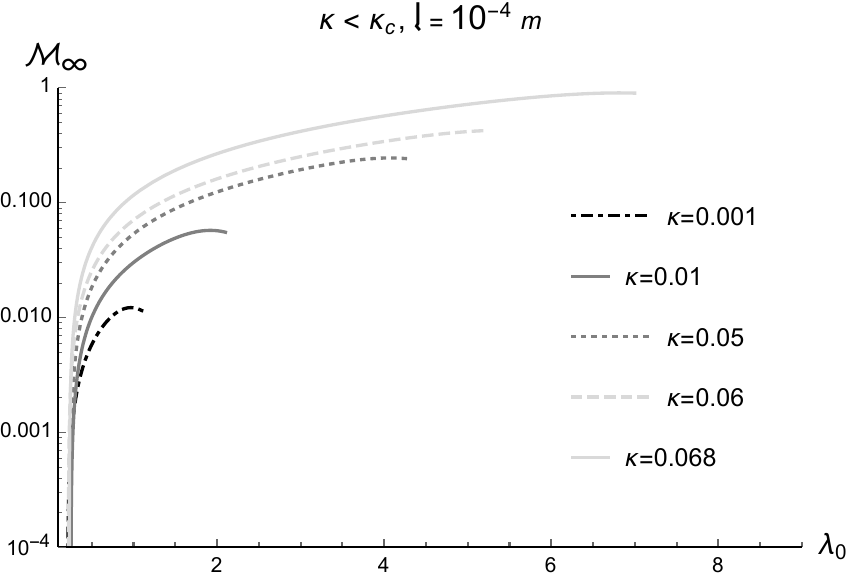}
\caption{
$\mathcal{M}_\infty$ as a function of the central field $\uplambda_{0}$, for $\mathfrak{l}=10^{-4}\,{\rm m}$, for $\kappa<\kappa_c$. The dot-dashed line regards $\kappa=0.001$, the grey line corresponds to  $\kappa=0.01$, the grey dotted line depicts $\kappa=0.05$, the light-grey dashed line illustrates the $\kappa=0.06$ case, whereas $\kappa=0.068$ is shown by the light-grey line.}
\label{mxx21}
\end{figure}

Figs. \ref{mxx19} -- \ref{mxx21} show the asymptotic value of the mass function $\mathcal{M}_\infty$, as a function of the central scalar field $\uplambda_{0}$, for several values of the fluid brane tension, with $\kappa<\kappa_c$. They are qualitatively similar to the GR limit (see, e. g., Fig. 4 of Ref. \cite{Zloshchastiev:2021ncg}), although in the MGD case here analyzed $\mathcal{M}_{\infty}$ increases around one order of magnitude slower, as a function of the scalar field $\uplambda_{0}$. Besides, the absolute values of $\mathcal{M}_{\infty}$, for each fixed $\kappa$, are slightly higher than the GR limit. Still regarding Figs. \ref{mxx19} -- \ref{mxx21}, it is also worth emphasizing that the higher the value of the brane tension, the higher the values of $\mathcal{M}_{\infty}$ are, for each fixed $\kappa$. One can observe that the increment of the brane tension increases the value of the mass function $\mathcal{M}_{\infty}$. The MGD case in Figs. \ref{mxx19} -- \ref{mxx21} illustrate another scenario. Figs. \ref{mxx19} and \ref{mxx20} respectively comprise the MGD 
cases $\mathfrak{l}=10^{-8}\,{\rm m}$ and $\mathfrak{l}=10^{-6}\,{\rm m}$ and for all values of $\kappa$, there are absolute maxima of $\mathcal{M}_{\infty}$ with null derivative. 
On the other hand, Fig. \ref{mxx21} differs from Figs. \ref{mxx19} and \ref{mxx20} by a fundamental aspect. Indeed, when $\mathfrak{l}=10^{-4}\,{\rm m}$, the function $\mathcal{M}_{\infty}$ has one absolute minimum when $\mathcal{M}_{\infty}=0$, and one absolute maximum with non-null derivative, corresponding to the endpoints of the $\uplambda_0$ range whereat $\mathcal{M}_{\infty}$ ceases to exist, for all values $\kappa\gtrapprox 0.054$. One can also verify numerically that in the brane tension range $\gamma\gtrapprox 4.623\times10^{-2} \;{\rm GeV}$, no maxima with null derivative exist for $\mathcal{M}_{\infty}$ other than the endpoints of the $\uplambda_0$ range whereat the values of $\mathcal{M}_{\infty}$ break off.

Table \ref{scalarmasses3} lists the 2-tuples corresponding to the central scalar field $\uplambda_{0}$ and the respective peaks of the $\mathcal{M}_{\infty}$ function, in Figs. \ref{mxx19} -- \ref{mxx21}. A first and important observation is that the central field $\uplambda_{0}$ strongly constrains $\mathcal{M}_{\infty}$, for each value of $\kappa$. Beyond this range of $\uplambda_{0}$ there is no real solutions for $\mathcal{M}_{\infty}$. Hence, for each value of $\kappa$, there is a maximum value of $\uplambda_{0}$ above which $\mathcal{M}_{\infty}$ ceases to exist. The suprema of the scalar field $\uplambda_{0}$, that represent a superior bound to the existence of $\mathcal{M}_{\infty}$, are presented in Table \ref{scalarmasses3}.

\begin{table}[h!]
\begin{center}
\medbreak
--------------\;\; 2-tuple $\left(\uplambda_{0}, \mathcal{M}_\infty\right)$, $\;\;\;\kappa<\kappa_c$ --------------\medbreak
\begin{tabular}{||c|c|c|c||}
\hline\hline& \;$\mathfrak{l}=10^{-8}\,{\rm m}$\; & \;$\mathfrak{l}=10^{-6}\,{\rm m}$ \;&\; $\mathfrak{l}=10^{-4}\,{\rm m}$\; \\
    \hline\hline
 $\kappa=0.001$&\; (1.412,\,0.0003)\;&\; (1.214,\,0.005)\;&\; (1.108,\,0.017) \;\\ \hline
 $\kappa=0.01$&\; (2.319,\,0.004)\;&\; (2.220.0.056)\;&\; (2.095,\,0.056) \;\\ \hline
$\kappa=0.05$&\; (4.776,\,0.077)\;&\; (4.776,\,0.293)\;&\; (4.311,\,0.213) \;\\\hline
$\kappa=0.06$&\; (5.910,\,0.231)\;&\; (5.611,\,0.312)\;&\; (5.207,\,0.378) \;\\\hline
$\kappa=0.068$&\; (9.031,\,0.418)\;&\; (7.502,\,0.763)\;&\; (7.009,\,0.810) \;\\\hline
\hline\hline
\end{tabular}
\caption{Suprema of the scaled scalar field $\uplambda_{0}$ (abscissa) and its corresponding $\mathcal{M}_{\infty}$ value (ordinate). Data from Figs. \ref{mxx19} -- \ref{mxx21}.} \label{scalarmasses3}
\end{center}
\end{table}
\noindent

\begin{figure}[h!]
\centering
\centering
	\includegraphics[width=8.2cm]{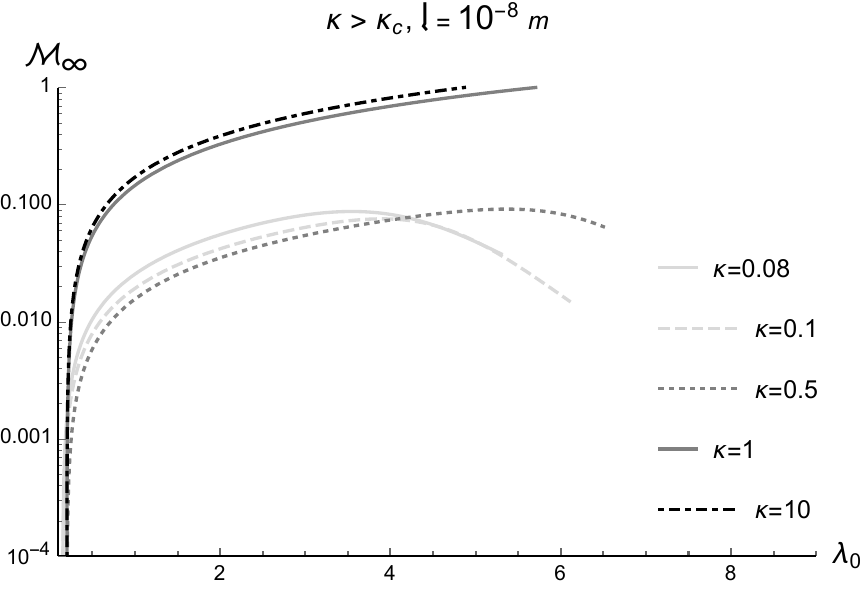}
\caption{
$\mathcal{M}_\infty$ as a function of the central field $\uplambda_{0}$, for $\mathfrak{l}=10^{-8}\,{\rm m}$, for $\kappa>\kappa_c$. The dot-dashed line regards $\kappa=0.08$, the grey line corresponds to  $\kappa=0.1$, the grey dotted line depicts $\kappa=0.5$, the light-grey dashed line illustrates the $\kappa=1$ case, whereas $\kappa=10$ is shown by the light-grey line.}
\label{mxx22}
\end{figure}

\begin{figure}[h!]
\centering
\centering
	\includegraphics[width=8.2cm]{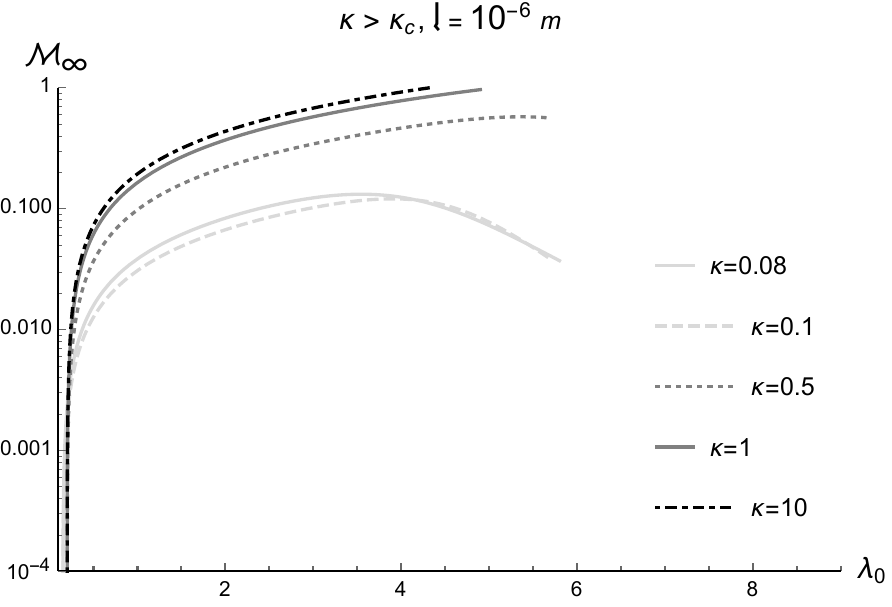}
\caption{$\mathcal{M}_\infty$ as a function of the central field $\uplambda_{0}$, for $\mathfrak{l}=10^{-6}\,{\rm m}$, for $\kappa>\kappa_c$. The light-grey line regards $\kappa=0.08$, the light-grey line corresponds to  $\kappa=0.1$, the grey dotted line depicts $\kappa=0.5$, the grey dashed line illustrates the $\kappa=1$ case, whereas $\kappa=10$ is shown by the dot-dashed line.}
\label{mxx23}
\end{figure}

\begin{figure}[h!]
\centering
\centering
	\includegraphics[width=8.2cm]{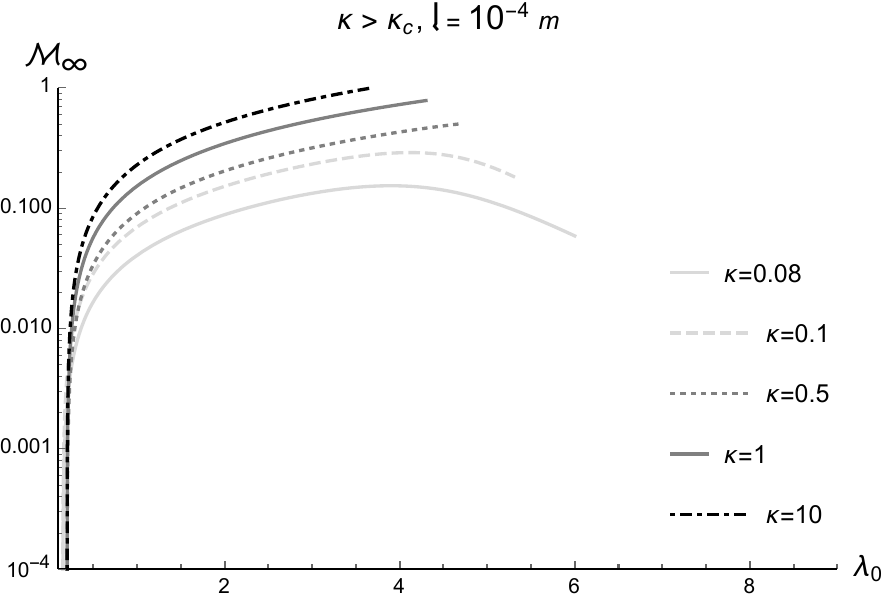}
\caption{
$\mathcal{M}_\infty$ as a function of the central field $\uplambda_{0}$, for $\mathfrak{l}=10^{-4}\,{\rm m}$, for $\kappa>\kappa_c$. The light-grey line regards $\kappa=0.08$, the light-grey line corresponds to  $\kappa=0.1$, the grey dotted line depicts $\kappa=0.5$, the grey dashed line illustrates the $\kappa=1$ case, whereas $\kappa=10$ is shown by the dot-dashed line.}
\label{mxx24}
\end{figure}

On the other hand, the case $\kappa>\kappa_c$ is shown in Figs. \ref{mxx22} -- \ref{mxx24} and maintain quite few qualitative similarities to the GR limit. In the MGD case here discussed the $\mathcal{M}_{\infty}$ increases slower, with respect to the scalar field $\uplambda_{0}$, than the GR limit. Also, the absolute values of $\mathcal{M}_{\infty}$, for each fixed $\kappa$, are slightly higher than the GR limit. The higher the value of the brane tension, the higher the values of $\mathcal{M}_{\infty}$ are, for each fixed $\kappa$. One can observe that the increment of the brane tension increases the value of the mass function $\mathcal{M}_{\infty}$. However, in the GR limit, $\mathcal{M}_{\infty}$ presents: a) for all values of $\kappa\lesssim0.078$, absolute maxima; b) for all values of $\kappa\gtrapprox0.078$, both absolute minima (besides the obvious minimum when $\mathcal{M}_{\infty}=0$) and absolute maxima. However, in the MGD case of Figs. \ref{mxx22} -- \ref{mxx24} the function $\mathcal{M}_{\infty}$ presents
one minimum $\mathcal{M}_{\infty}=0$ and, for a more precise analysis, one must split into three relevant cases. In the first one, when $\mathfrak{l}=10^{-8}\,{\rm m}$ (Fig. \ref{mxx22}), the function $\mathcal{M}_{\infty}$ presents one absolute maximum with null derivative, regarding the endpoints of the $\uplambda_0$ range at which $\mathcal{M}_{\infty}$ breaks off, for all values $\kappa\gtrapprox 0.819$. This behavior changes for $\kappa= 0.819$, when for all values $\kappa\lessapprox 0.819$ there is one absolute maximum of $\mathcal{M}_{\infty}$ with null derivative. For the second case, where $\mathfrak{l}=10^{-6}\,{\rm m}$ corresponds to Fig. \ref{mxx23}, the function $\mathcal{M}_{\infty}$ has one absolute maximum with non-null derivative, corresponding to the endpoints of the $\uplambda_0$ range at which $\mathcal{M}_{\infty}$ ceases to exist, for all values $\kappa\gtrapprox 0.380$ and in the range $\kappa\lessapprox 0.380$ the function $\mathcal{M}_{\infty}$ exhibits one absolute maximum with null derivative. Finally, for a higher value of the brane tension, the third case consists of $\mathfrak{l}=10^{-4}\,{\rm m}$ (Fig. \ref{mxx24}). The function $\mathcal{M}_{\infty}$ presents
one absolute maximum with non-null derivative for $\kappa\gtrapprox 0.514$. In the range $\kappa\lessapprox 0.514$, there is one absolute maximum of $\mathcal{M}_{\infty}$ with null derivative. 

Table \ref{scalarmasses4} shows the 2-tuples corresponding to the central scalar field $\uplambda_{0}$ and the respective maxima of the $\mathcal{M}_{\infty}$ function, in Figs. \ref{mxx22} -- \ref{mxx24}. Again, the central field $\uplambda_{0}$ also constrains the asymptotic value of the mass function $\mathcal{M}_{\infty}$, for each value of $\kappa$. Far away this range of $\uplambda_{0}$, no real solutions for $\mathcal{M}_{\infty}$ exist. Consequently, for each value of $\kappa$, there is a maximum value of $\uplambda_{0}$ above which $\mathcal{M}_{\infty}$ breaks off. The suprema of the scaled scalar field $\uplambda_{0}$ characterizing a superior limiting bound to $\mathcal{M}_{\infty}$ are presented as abciss\ae\; of the 2-tuples, in Table \ref{scalarmasses4}.

\begin{table}[H]
\begin{center}
\medbreak
--------------\;\; 2-tuple $\left(\uplambda_{0}, \mathcal{M}_\infty\right)$, $\;\;\;\kappa>\kappa_c$ --------------\medbreak
\begin{tabular}{||c|c|c|c||}
\hline\hline& \;$\mathfrak{l}=10^{-8}\,{\rm m}$\; & \;$\mathfrak{l}=10^{-6}\,{\rm m}$ \;&\; $\mathfrak{l}=10^{-4}\,{\rm m}$\; \\
    \hline\hline
 $\kappa=0.08$&\; (5.324,0.023)\;&\; (5.812,0.020)\;&\; (6.002,0.061) \;\\ \hline
 $\kappa=0.1$&\; (6.101,0.018)\;&\; (5.739.0.021)\;&\; (5.291,0.198) \;\\ \hline
$\kappa=0.5$&\; (6.520,0.038)\;&\; (5.703,0.478)\;&\; (4.696,0.511) \;\\\hline
$\kappa=1$&\; (7.892,0.998)\;&\; (4.981,0.923)\;&\; (4.306,0.897) \;\\\hline
$\kappa=10$&\; (8.003,0.994)\;&\; (4.500,0.996)\;&\; (4.053,0.997) \;\\\hline
\hline\hline
\end{tabular}
\caption{Suprema of the scaled scalar field $\uplambda_{0}$ (abscissa) and its corresponding $\mathcal{M}_{\infty}$ value (ordinate). Data from Figs. \ref{mxx22} -- \ref{mxx24}.} \label{scalarmasses4}
\end{center}
\end{table}

\blt{In Figs. \ref{mxx26} -- \ref{mxx28}, the asymptotic value of the mass function $\mathcal{M}_\infty$ is displayed as a function of the squared frequency $\Upomega^2$, for values of $\kappa$ above the critical value $\kappa_c$.
\begin{figure}[h]
\centering
\centering
	\includegraphics[width=7.5cm]{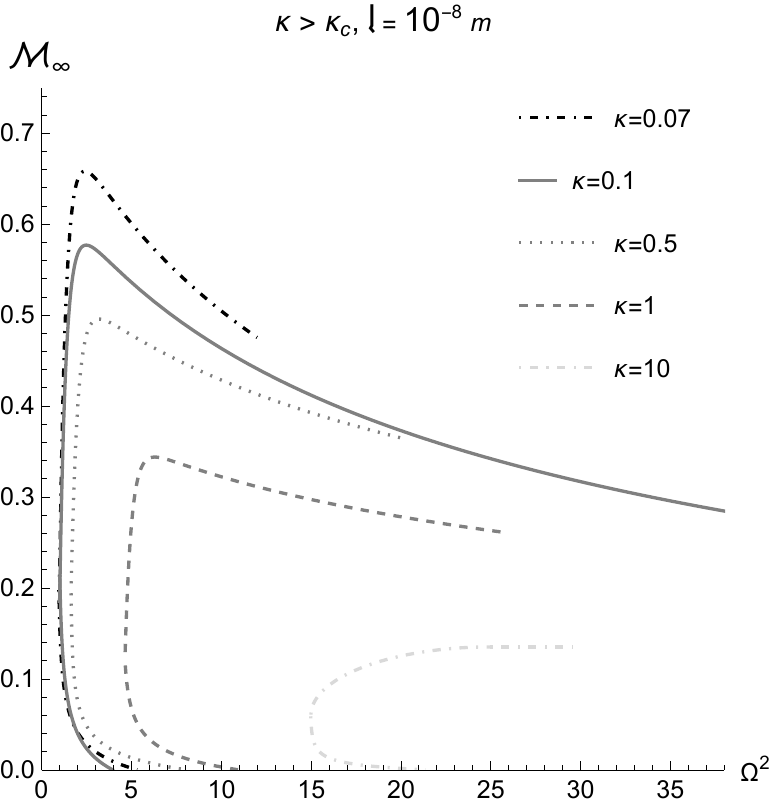}
\caption{\blt{ Plot of $\mathcal{M}_\infty$ as a function of $\Upomega^2$, for $\mathfrak{l}=10^{-8}\,{\rm m}$ and $\kappa>\kappa_c$. 
The black dot-dashed line regards $\kappa=0.07$, the dark grey line corresponds to  $\kappa=0.1$, the dark grey dotted line depicts $\kappa=0.5$, the grey dashed line illustrates the $\kappa=1$ case, whereas $\kappa=10$ is shown by the light-grey dot-dashed line.}}
\label{mxx26}
\end{figure}
\begin{figure}[H]
\centering
\centering
	\includegraphics[width=6.5cm]{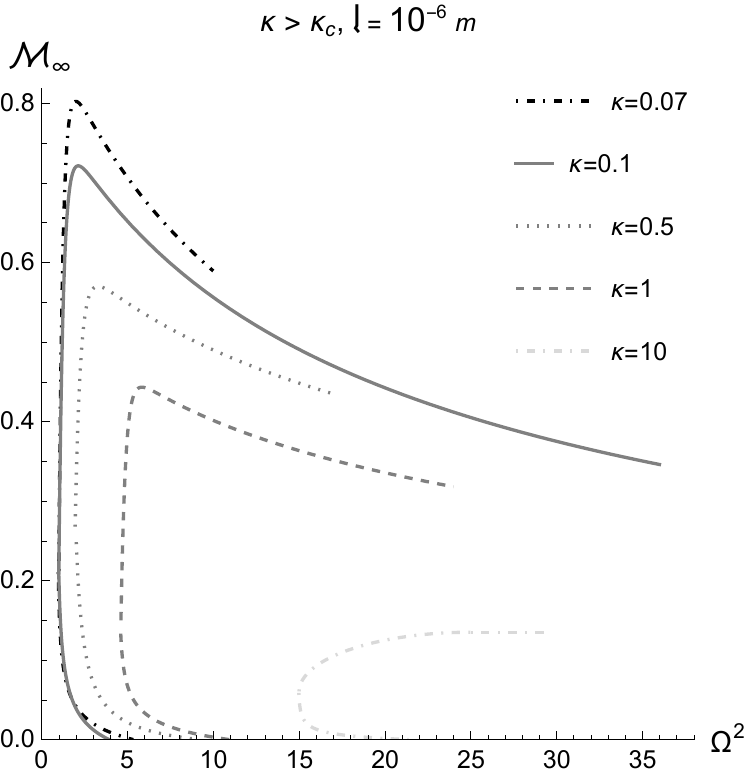}
\caption{\blt{ Plot of $\mathcal{M}_\infty$ as a function of $\Upomega^2$, for $\mathfrak{l}=10^{-6}\,{\rm m}$ and $\kappa>\kappa_c$. 
The black dot-dashed line regards $\kappa=0.07$, the dark grey line corresponds to  $\kappa=0.1$, the dark grey dotted line depicts $\kappa=0.5$, the grey dashed line illustrates the $\kappa=1$ case, whereas $\kappa=10$ is shown by the light-grey dot-dashed line.}}
\label{mxx27}
\end{figure}
\begin{figure}[H]
\centering
\centering
	\includegraphics[width=6.5cm]{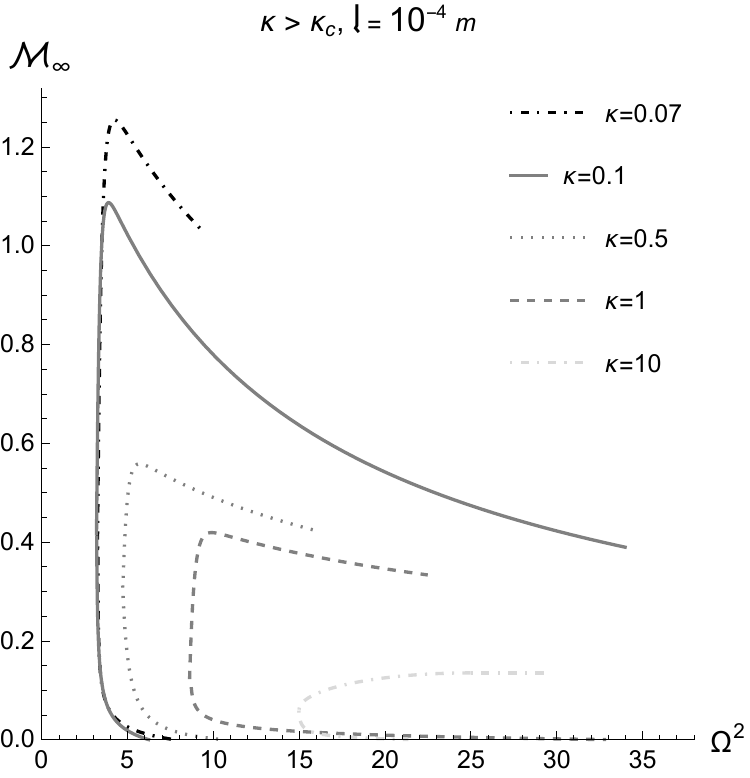}
\caption{\blt{ Plot of $\mathcal{M}_\infty$ as a function of $\Upomega^2$, for $\mathfrak{l}=10^{-4}\,{\rm m}$ and $\kappa>\kappa_c$. 
The black dot-dashed line regards $\kappa=0.07$, the dark grey line corresponds to  $\kappa=0.1$, the dark grey dotted line depicts $\kappa=0.5$, the grey dashed line illustrates the $\kappa=1$ case, whereas $\kappa=10$ is shown by the light-grey dot-dashed line.}}
\label{mxx28}
\end{figure}
\noindent Although the graphics in Figs. \ref{mxx26} -- \ref{mxx28} keep a relative qualitative similarity to the GR case, the role played by the brane tension-dependent MGD parameter (\ref{Lk}) makes profound quantitative differences.  For each fixed value of $\kappa$, the higher the brane tension, the lower the peak of the asymptotic value of the mass function $\mathcal{M}_\infty$ is, as a function of the squared frequency $\Upomega^2$. For fixed values of the fluid brane tension, the higher the value of $\kappa$, the lower the peak of the asymptotic value of the asymptotic mass function $\mathcal{M}_\infty$ is. Besides, an important feature resides in the fact that there is a maximal frequency, for each value of $\kappa$, above which $\mathcal{M}_\infty$ ceases to exist. 
\begin{table}[H]
\begin{center}
\medbreak
\begin{tabular}{||c|c|c|c||}
\hline\hline& \;$\mathfrak{l}=10^{-8}\,{\rm m}$\; & \;$\mathfrak{l}=10^{-6}\,{\rm m}$ \;&\; $\mathfrak{l}=10^{-4}\,{\rm m}$\; \\
    \hline\hline
 $\kappa=0.07$&\; (12.10, 0.478)\;&\; (5.812, 0.592)\;&\; (9.34, 1.046) \;\\ \hline
 $\kappa=0.1$&\; (38.07, 0.300)\;&\; (36.51, 0.379)\;&\; (34.20, 0.398) \;\\ \hline
$\kappa=0.5$&\; (19.88, 0.359)\;&\; (16.09, 0.427)\;&\; (17.29, 0.430) \;\\\hline
$\kappa=1$&\; (26.05, 0.267)\;&\; (23.81, 0.316)\;&\; (22.58, 0.349) \;\\\hline
$\kappa=10$&\; (29.61, 0.135)\;&\; (30.67, 0.198)\;&\; (28.13, 0.279) \;\\\hline
\hline\hline
\end{tabular}
\caption{Maximal range of the squared frequency, $\Upomega^2$ (abscissa), and the correspondent value of $\mathcal{M}_\infty$ (ordinate). Data from Figs. \ref{mxx26} -- \ref{mxx28}.} \label{scalarmasses14}
\end{center}
\end{table}}

\begin{figure}[H]
\centering
\centering
	\includegraphics[width=7cm]{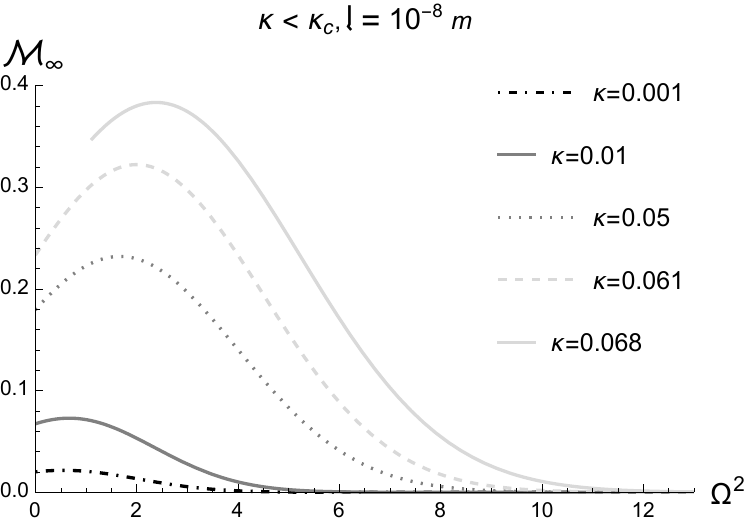}
\caption{\blt{ Plot of $\mathcal{M}_\infty$ as a function of $\Upomega^2$, for $\mathfrak{l}=10^{-8}\,{\rm m}$ and $\kappa<\kappa_c$. 
The light-grey line regards $\kappa=0.001$, the light-grey dashed line corresponds to  $\kappa=0.01$, the black dotted line depicts $\kappa=0.05$, the dark grey line illustrates the $\kappa=0.061$ case, whereas $\kappa=0.68$ is shown by the black dot-dashed line.}}
\label{mxx29}
\end{figure}

\begin{figure}[H]
\centering
\centering
	\includegraphics[width=7cm]{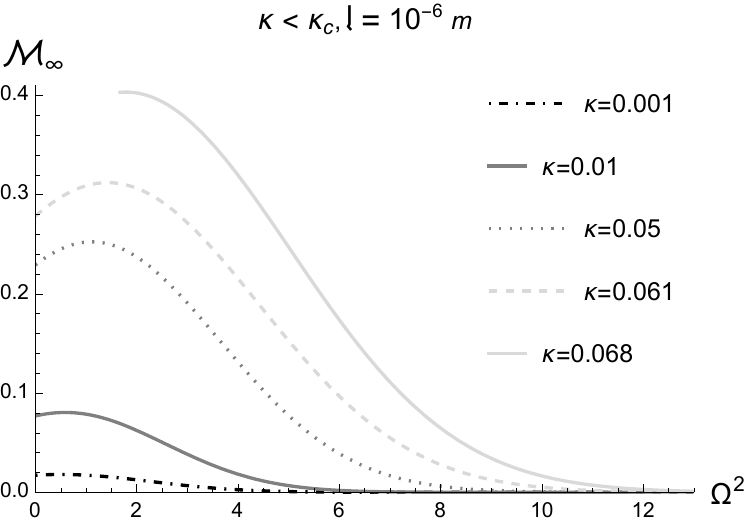}
\caption{\blt{ Plot of $\mathcal{M}_\infty$ as a function of $\Upomega^2$, for $\mathfrak{l}=10^{-6}\,{\rm m}$ and $\kappa<\kappa_c$. 
The light-grey line regards $\kappa=0.001$, the light-grey dashed line corresponds to  $\kappa=0.01$, the black dotted line depicts $\kappa=0.05$, the dark grey line illustrates the $\kappa=0.061$ case, whereas $\kappa=0.68$ is shown by the black dot-dashed line.}}
\label{mxx30}
\end{figure}

\begin{figure}[h]
\centering
\centering
	\includegraphics[width=7cm]{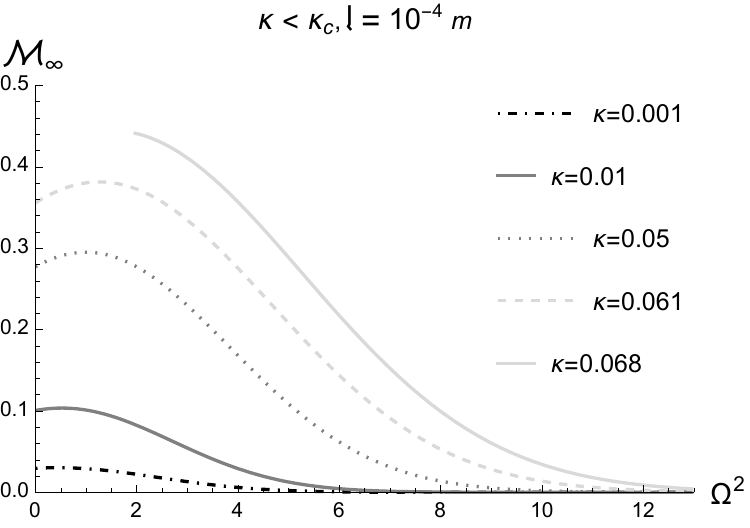}
\caption{\blt{ Plot of $\mathcal{M}_\infty$ as a function of $\Upomega^2$, for $\mathfrak{l}=10^{-4}\,{\rm m}$ and $\kappa<\kappa_c$. 
The light-grey line regards $\kappa=0.001$, the light-grey dashed line corresponds to  $\kappa=0.01$, the black dotted line depicts $\kappa=0.05$, the dark grey line illustrates the $\kappa=0.061$ case, whereas $\kappa=0.68$ is shown by the black dot-dashed line.}}
\label{mxx31}
\end{figure}

\blt{Now, Figs. \ref{mxx29} -- \ref{mxx31} display the asymptotic value of the mass function $\mathcal{M}_\infty$, as a function of the squared frequency $\Upomega^2$, for values of $\kappa$ below the critical value $\kappa_c$. Again these graphics have a relative qualitative analogy to the GR case, but quantitative aspects make them quite different, due to the finite brane tension in the MGD solutions.  For each fixed value of $\kappa$, the higher the brane tension, the lower the peak of $\mathcal{M}_\infty$ is, as a function of the squared frequency $\Upomega^2$. Contrary to the case when $\kappa>\kappa_c$, for fixed values of the brane tension, the higher the value of $\kappa$, the higher the peak of the asymptotic value of $\mathcal{M}_\infty$ is. Just for values close to the critical $\kappa_c$, there are non-null values of the frequency which correspond to a minimal value of $\mathcal{M}_\infty$, in Figs. \ref{mxx29} -- \ref{mxx31}, illustrated for $\kappa=0.068$. Below these values of the frequency, $\mathcal{M}_\infty$ ceases to exist. }

To end this section, let us provide a concrete estimative of MGD-decoupling modifications with respect to GR. Ref. \cite{Zloshchastiev:2021ncg} showed that for Solar size astrophysical stellar objects with mass $M_\odot\sim 1.99\times 10^{30}$ kg in the GR regime, a Schwarzschild superfluid star has radius $
R\sim1/\sqrt{a_{1\odot}} 
\approxeq 16.92 \;\text{km},
$ 
where 
$a_{1\odot}=
2.86 \times 10^{-8}\;{\rm m}^{-2}$.  
Here, using Eq. (\ref{eins15}), MGD superfluid stars have radii 
that are slightly altered by a term involving the parameter $\alpha$ that governs the source term of MGD-decoupling (Eq. (\ref{eins7})). Besides, the Einstein--Klein--Gordon system involves the mass function \eqref{eoms14}, which is also function of the brane tension and the coupling parameter $a_1$ as well, as it includes terms that depends on the MGD metric coefficients $a(r)$ and $b(r)$. Assuming the physically realistic values for the finite fluid brane tension value, 
$\gamma \sim 3\times10^{-6} \;{\rm GeV}$, and for the MGD parameter, $\mathfrak{l}=10^{-4}$ m, MGD superfluid stars have effective radius $
R \approxeq 17.01 \;\text{km}.
$ It is a considerable difference of $0.53\%$, susceptible to observation. When $\mathfrak{l}=10^{-8}$ m, one has $
R \approxeq 16.95 \;\text{km},
$ a difference of $0.23\%$ regarding the GR case. 

\section{Conclusions}
\label{4}
Some results regarding superfluid stars in the GR limit \cite{Zloshchastiev:2021ncg} are qualitatively similar to the ones obtained in this work, scrutinizing MGD superfluid stars. However, most of our results show that the finite fluid brane tension and the MGD parameter bring new results that comply with a more realistic and physically feasible model, whose eventual gravitational-wave observations can be more reliable in a finite brane tension setup. 
The maximal masses of the MGD superfluid stars increase at a slower pace, with respect to the minimally-coupled scalar field, when compared to the GR                                                                                                                                                                                                                                                                                                                                                                                                                                                                                                                                                                                                                                                                                                                                                                                  limit. Besides, higher values of the brane tension can increase the effective range of the scalar field that satisfies the Schr\"odinger-like equation with logarithmic potential, coupled to gravity. It illustrates the effect of a finite brane tension, that attenuates the dependence of the MGD superfluid stars with the scalar field.
On the other hand, the results analyzed show that the absolute values of the mass function, for each fixed value of the coupling parameter $\kappa$, are slightly higher than the GR limit. This time, the finite fluid brane tension plays the role of heightening the maximal values of the mass function, $\mathcal{M}_{\infty}$ are, for each fixed $\kappa$. This difference may be further probed by observations of precise signatures of the MGD-decoupling. 
MGD superfluid stars, driven by self-interacting logarithmic scalar fields, can generate regular equilibria stellar configurations. 
This setup can be employed to report astrophysical MGD stellar distributions. The study here completely differs from the Einstein--Klein--Gordon system studied in Ref. \cite{Ovalle:2018ans}, which derives self-gravitating compact stellar systems surrounded by a coupled scalar field. The scalar field in this context defined a fluid that dilutes the gravitational field at the nuclear source of the Schwarzschild-like metric. Here, the scalar field satisfies the covariant equation (\ref{KGeuler}), that mimics a Schr\"odinger-like equation with logarithmic potential, that precisely models MGD superfluid stars.

\subsection*{Acknowledgements}

RdR~is grateful to FAPESP (Grants No. 2021/01089-1 and No. 2017/18897-8) and the National Council for Scientific and Technological Development -- CNPq (Grants No. 303390/2019-0 and No. 406134/2018-9), for partial financial support.

\bibliography{bib_DSS}

\end{document}